\pdfoutput=1
\documentclass{aastex61}

\newcommand{\beq}{\begin{equation}}         
\newcommand{\eeq}{\end{equation}}           
\newcommand{\bfi}{\begin{figure}}           
\newcommand{\efi}{\end{figure}}             

\shortauthors{Ledvina, Heyrovsk\'y, \& Dov\v{c}iak}

\shorttitle{Quasar X-ray Line Microlensing}

\begin{document}

\title{X-RAY LINE PROFILE VARIATIONS DURING QUASAR MICROLENSING}

\author{Luk\'a\v{s} Ledvina}
\affil{Institute of Theoretical Physics, Faculty of Mathematics and Physics, Charles University, V Hole\v{s}ovi\v{c}k\'ach 2, 18000~Praha, Czech Republic}
\author{David Heyrovsk\'y}
\affil{Institute of Theoretical Physics, Faculty of Mathematics and Physics, Charles University, V Hole\v{s}ovi\v{c}k\'ach 2, 18000~Praha, Czech Republic}
\author{Michal Dov\v{c}iak}
\affil{Astronomical Institute, Academy of Sciences, Bo\v{c}n\'i II 1401, 14131 Praha, Czech Republic}
\email{ledvina@utf.mff.cuni.cz, david.heyrovsky@mff.cuni.cz, dovciak@asu.cas.cz}

\begin{abstract}

Observations of several gravitationally microlensed quasars in X-rays revealed variations in the profile of the iron K$\alpha$ line in the course of microlensing events. We explore the effect by simulating a microlensing caustic crossing a spatially resolved model of emission from a thin accretion disk around a Kerr black hole. We demonstrate the sequence of spectral changes during the event, in particular the appearance of additional peaks and edges in the line profile due to microlensing. We trace the origin of these features to points on the disk, at which the total energy shift ($g$-factor) contours are tangent to the caustic. Contours tangent from the inner side of the caustic generate peaks, while those tangent from its outer side generate edges. We derive analytical shapes of the generated features and map the peak strength as a function of position of the tangent point on the disk. Since the features are determined by the positional geometry of the caustic relative to the $g$-factor contours, the same type of behavior can be expected in a much broader range of emission models. The sequence of line profile changes thus serves as a sensitive probe of the geometry and physics of the innermost region of the quasar accretion disk.

\end{abstract}


\section{INTRODUCTION}
\label{sec:intro}

Quasar microlensing occurs as a secondary effect in quasar images that are formed by gravitational lensing sufficiently close to the lens galaxy, so that they are superimposed over its stellar population. Proper motion of the quasar relative to the stars then leads to variations in gravitational amplification of the image flux, with peaks occurring whenever the quasar crosses the microlensing caustic network produced by the stars' gravitational field \citep[for a detailed review see][]{schmidt_etal10}.

During the crossing, the high angular resolution of the caustic provides a unique opportunity to spatially resolve the emitting region of the quasar. Since the extent of the region varies with wavelength, observations of quasar microlensing in multiple wavelengths can be used to test the spatial structure of quasar emission models.

The sensitivity of microlensing light curves to the structure of the emitting region was demonstrated first in simulations \citep[e.g.,][]{jaroszynski_etal92,agol_etal99}. Most observations of quasar microlensing are based on optical and UV photometry. Such data have been used to measure the size of the emitting region as a function of wavelength \citep[e.g.,][]{eigenbrod_etal08}, or to measure the temperature profile of the quasar accretion disk \citep{munoz_etal16}. A number of studies have compared optical and X-ray observations to demonstrate the compact nature of the X-ray-emitting region \citep{dai_etal10,morgan_etal12,jimenez-vicente_etal15,macleod_etal15}. Generally, the smaller size of the emitting region implies a higher microlensing variability, since the amplification is averaged over a smaller region of the caustic network.

More advanced models were later used to invert observations of individual caustic-crossing events \citep{abolmasov_shakura12,mediavilla_etal15}. The inversion of optical light curves is affected by the relatively large size of the emitting region. As a result, one needs to integrate over larger areas of simulated caustic networks; a simple local linear fold caustic approximation is rarely sufficient. The problem is reduced when studying X-ray microlensing, due to the smaller emitting region of the quasar. \cite{yonehara_etal98} demonstrated the possibility of resolving the innermost accretion disk by X-ray observations of quasar microlensing. By now X-ray microlensing has been convincingly detected in a growing list of systems \citep{chartas_etal02,chartas_etal09,zimmer_etal11,guerras_etal17}. The \emph{Chandra X-ray Observatory} has been particularly useful, primarily due to its high angular resolution, which permits separate observations of individual macroimages of the quasar. As shown by \cite{chen_etal13b}, analysis of such observations requires simulations involving strong-gravity effects, since the radiation is emitted in the vicinity of the quasar black hole.

In addition to the photometric observations discussed above, X-ray spectroscopy provides further interesting possibilities. In particular, many lensed quasars exhibit a prominent iron K$\alpha$ line \citep{walton_etal15} originating from the innermost disk. The shape of its profile is dominated by strong-gravity effects \citep{laor91,jovanovic12}, including gravitational redshift, light bending, Doppler broadening, and beaming due to orbital motion close to the black hole.
A caustic crossing the innermost disk would differentially amplify regions generating different shapes of the profile, leading to variations in the observed integrated line profile.

Changes in the observed Fe K$\alpha$ line profile during a microlensing event were first reported by \cite{chartas_etal12} for quasar RX J1131--1231. Although only a few spectra were measured during the crossing, the line intriguingly changed from a single-peaked to a double-peaked profile and back again. More recently, line profile changes have been reported in two other quasars, QJ 0158–4325 and SDSS 1004+4112 \citep{chartas_etal17}.

Explaining such line profile variations and interpreting them in terms of the properties of the innermost region of the quasar requires both advanced modeling of emission in the strong-gravity regime and accounting for the origin of microlensing-induced features. Nevertheless, already the simplest weak-gravity model of a line emitted from a Keplerian disk microlensed by a single star \citep{heyrovsky_etal97} showed that the lens generates an additional peak in the line profile. The peak position corresponds to the velocity in the disk at the point directly behind the lens, which achieves maximum amplification.

Gradually more realistic and more advanced simulations followed: a relativistic disk model with a single lens \citep{popovic_etal01}, an accretion disk around a Kerr or Schwarzschild black hole with a single lens or a linear caustic \citep{popovic_etal03a,popovic_etal03b}, a similar disk lensed by a microlensing caustic network \citep{popovic_etal06}, and a disk with an additional absorbing region \citep{jovanovic_etal09}.

\citet{neronov_etal16} studied the evolution of the line profile from a microlensed thin disk around a Kerr black hole and pointed out the generation and progression of red and blue ``edges'' of the line. \cite{krawczynski_chartas17} used a similar model and showed that microlensing may generate line profile variations similar to the \citet{chartas_etal12} observations.

The main goal of our work was to explore the character and origin of microlensing-induced features in the line profile. Initial results of our simulations \citep{ledvina_heyrovsky15} already showed the generation of peaks and edges, which we set out to study in detail in the present paper.

In \S~\ref{sec:simulate} we describe our simulations, including the disk emission model, microlensing caustic model, and the computation of the microlensed spectrum. The computed line profile variations are described in \S~\ref{sec:spExample}, illustrating the gradual formation and progression of line profile features in the course of a caustic crossing. The evolution of peaks, edges, and higher-order spectral features is shown in detail, and their connection to local properties of the disk is pointed out. In \S~\ref{sec:analytical} we prove this connection analytically by using a local model of disk emission, which enables us to derive the analytical shapes of the generated peaks and edges. We map the peak strength as a function of caustic position on the disk and demonstrate the accuracy of the analytical peak model by fitting computed microlensed line profiles. We discuss the implications, limitations, and possible extensions of our work in \S~\ref{sec:discussion}.

\section{SIMULATING X-RAY LINE MICROLENSING}
\label{sec:simulate}

\subsection{Quasar Accretion Disk Emission Model}
\label{sec:emission}

For illustration we use a simple model of the X-ray-emitting region of the quasar, consisting of a thin accretion disk in the equatorial plane of a Kerr black hole. We concentrate on the emission in the spectral vicinity of a single emission line, exemplified here by the typically most prominent iron K$\alpha$ line. At a given point of the optically thick disk we describe the local rest-frame emission by a narrow line superimposed on a power-law continuum. We model the specific photon intensity of the continuum in the rest frame of an emitting point by
\beq
I_{\rm em}^{\rm cont}(E_{\rm em};r) = I_0^{\rm cont}\,r^{-q}\,E_{\rm em}^{-\Gamma}\,,
\label{eq:intensity-cont-em}
\eeq
where $E_{\rm em}$ is the photon energy in the emission-point rest frame, $\Gamma$ is the spectral power-law index, $r$ is the radial Boyer--Lindquist coordinate of the emitting point, and $I_0^{\rm cont}$ is a constant. We assume a power-law decline of the emissivity with radius, parameterized by a radial decline index $q$ (assumed energy independent).

The line width in the emission-point rest frame is negligible in comparison with the disk-integrated line width, which is dominated by gravitational redshift and the relativistic Doppler effect. Hence, we can safely model the iron line specific photon intensity in the emission-point rest frame by a delta-function profile,
\beq
I_{\rm em}^{\rm Fe}(E_{\rm em};r) = I_0^{\rm Fe}\,r^{-q}\,\delta\left[E_{\rm em}-E_{\rm Fe}\right]\,,
\label{eq:intensity-iron-em}
\eeq
where the rest-frame energy of the iron K$\alpha$ line $E_{\rm Fe}=$ 6.4 keV and $I_0^{\rm Fe}$ is a constant\footnote{Note that the ratio of the constants $I_0^{\rm Fe}/I_0^{\rm cont}$ has the physical dimension of $E_{\rm em}^{-\Gamma+1}$.} . For simplicity we assume the same radial decline index $q$ as for the continuum.

Next, we need to transform the emission-point specific intensities $I_{\rm em}(E_{\rm em};r)$ given by Equations~(\ref{eq:intensity-cont-em}) and (\ref{eq:intensity-iron-em}) to the rest frame of an asymptotically distant observer viewing the disk under an inclination angle $i$. In order to connect a point $(\alpha,\beta)$ in the plane of the observer's sky to a corresponding emission point on the disk, we integrate backward along photon geodesics in the Kerr metric following \mbox{\citet{dovciak_etal04}}. In this way we obtain numerically the Boyer--Lindquist emission radius $r(\alpha,\beta)$. The found emission point and its orbital velocity define the total energy shift (Doppler and gravitational) for a photon following the geodesic from the disk to the observer. The shift is described by the dimensionless $g$-factor,
\beq
g(\alpha,\beta) = E / E_{\rm em},
\label{eq:g-factor}
\eeq
where $E$ is the photon energy in the rest frame of the observer. Note that $r(\alpha,\beta)$ and $g(\alpha,\beta)$ depend also on the disk inclination $i$ and the black hole spin $a$. The specific photon intensity in the observer frame is then given by
\beq
I_{\rm obs}(E;\alpha,\beta) = g^2(\alpha,\beta)\,I_{\rm em}[\,E/g(\alpha,\beta);r(\alpha,\beta)\,]\,,
\label{eq:intensity-obs}
\eeq
where we utilized the invariance $I_{\rm obs}/E^2=I_{\rm em}/E^2_{\rm em}$ and Equation~(\ref{eq:g-factor}).

Observer-frame emission maps for the continuum and the line can be obtained by substituting for $I_{\rm em}$ from Equation~(\ref{eq:intensity-cont-em}) and (\ref{eq:intensity-iron-em}), respectively. For the continuum we get
\beq
I_{\rm obs}^{\rm cont}(E;\alpha,\beta) = I_0^{\rm cont} g^{\Gamma+2}(\alpha,\beta)\, r^{-q}(\alpha,\beta)\, E^{-\Gamma}\,,
\label{eq:intensity-cont-obs}
\eeq
i.e., the same power-law spectrum from any point $(\alpha,\beta)$, with amplitude spatially modulated by a combination of powers of the $g$-factor and the disk radial coordinate. For the iron line we get
\beq
I_{\rm obs}^{\rm Fe}(E;\alpha,\beta) = I_0^{\rm Fe} g^3(\alpha,\beta)\,r^{-q}(\alpha,\beta)\, \delta\left[E-E_{\rm Fe}\,g(\alpha,\beta)\,\right]\,,
\label{eq:intensity-iron-obs}
\eeq
a delta function from any point, shifted by the local $g$-factor, and with amplitude spatially modulated by a combination of powers of $g$ and $r$. The extra power of the $g$-factor comes from rescaling the argument of the delta function.

For illustration we present in Figure~\ref{fig:spec-int} sample observer-frame plane-of-the-sky maps of the iron line specific intensity $I_{\rm obs}^{\rm Fe}(\alpha,\beta)$. Both panels are computed for black hole spin $a=1$ (corresponding to maximal rotation) and radial index $q=3$. The counterclockwise rotating disk is viewed face-on ($i=0^\circ$) in the left panel and under inclination $i=70^\circ$ in the right panel. The plots are centered on the black hole; the coordinates are marked in units of the gravitational radius $r_{\rm g}=GM/c^2$, where $M$ is the black hole mass. In the central region the disk is truncated at the innermost stable circular orbit (ISCO), which is located at the $r=r_{\rm g}$ horizon in the maximally rotating case. The color scale corresponds to the amplitude-modulating factor preceding the delta function in Equation~(\ref{eq:intensity-iron-obs}), with values given in units of $I_0^{\rm Fe}\,r_{\rm g}^{-q}$. The plotted $g$-factor contours illustrate the relative shift of the delta-function line at the given position.

In the symmetric face-on case in the left panel of Figure~\ref{fig:spec-int} all contours are redshifted ($g<1$), with $g\to1$ far from the origin and $g=0$ at the black hole horizon. The radius of the central gap is larger than $r_{\rm g}$ owing to the geodesic-focusing effect of the black hole. As indicated by the color scale, the intensity drops rapidly outward and increases toward the center, with a sharp drop to zero at the horizon due to its $g$-factor dependence.

In the inclined $i=70^\circ$ case in the right panel, the $g=1$ no-shift contour is plotted in bold, with blue-shifted $g>1$ contours to its left and redshifted $g<1$ to its right. In the presented example the highest $g=1.304$ blue-shift comes from a point to the left of the hole, while the extreme $g=0$ redshift comes from the horizon. Far from the central region the contours converge to the familiar ``dipole'' pattern of an inclined Keplerian disk. The radial decline of the intensity is distorted by its $g$-factor dependence. The intensity peaks just to the left of the horizon and drops to zero along the horizon.

\subsection{Microlensing Caustic Model}
\label{sec:caustic}

In quasar microlensing the gravitational field of the local distribution of stars and continuous matter amplifies light passing through a particular region of the lens galaxy. The result is usually presented in the form of a map of the amplification of flux from a background point-like source as a function of its plane-of-the-sky position. A cut through the map along a given source trajectory yields the corresponding point-source light curve. An amplification map typically consists of areas with low amplification variation divided by a network of caustics, along which the amplification sharply diverges. The local structure of the caustics can be typically described by linear folds that connect sharply at cusps.

If the projected size of the source is smaller than the structure scale of the map, it is advantageous to use a simple local analytical approximation of the amplification instead of the full numerically computed map. This approach is justified in our case, since the X-ray-emitting region is limited to the innermost parts of the quasar accretion disk. We simulate a typical microlensing event by the crossing of a straight-line fold caustic, for which the point-source amplification
\beq
A(d_\perp) = A_0 + H[d_\perp]\,\sqrt{\frac{d_0}{d_\perp}}
\label{eq:fold}
\eeq
depends only on the coordinate $d_\perp$ perpendicular to the caustic \citep{schneider_etal92}. Here the parameter $d_0>0$ defines the strength of the caustic, the Heaviside function $H[d_\perp\geq0]=1$ and $H[d_\perp<0]=0$, and $A_0$ is the baseline amplification. For points outside the caustic $d_\perp<0$ and the amplification is constant, while for points inside the caustic $d_\perp>0$ and the amplification diverges as the inverse square root of the separation from the caustic.

For a caustic oriented so that the direction pointing inside subtends an angle $\psi$ from the $\alpha$-axis we can write
\beq
d_\perp(\alpha,\beta;\alpha_0,\beta_0)=(\alpha-\alpha_0)\cos{\psi}+(\beta-\beta_0)\sin{\psi}\,
\label{eq:perp-origin}
\eeq
where $(\alpha_0,\beta_0)$ are the coordinates of an arbitrary point on the caustic. For a caustic crossing the disk we can replace these by the velocity $v_\perp$ of its perpendicular motion and time $t$. We define the former in terms of the time $t_{\rm g}$ it takes the caustic to perpendicularly cross a distance corresponding to the gravitational radius $r_{\rm g}$,
\beq
v_\perp=\frac{r_{\rm g}}{t_{\rm g}}\,\sigma_{\rm enter}\,,
\label{eq:vel-perp}
\eeq
where $\sigma_{\rm enter}=1$ if the disk is entering the caustic (inside at $t\to\infty$) and $\sigma_{\rm enter}=-1$ if the disk is exiting the caustic (outside at $t\to\infty$). The perpendicular coordinate can then be computed from
\beq
d_\perp(\alpha,\beta;t)=\alpha\cos{\psi}+\beta\sin{\psi}+\frac{t}{t_{\rm g}}\,r_{\rm g}\,\sigma_{\rm enter}\,,
\label{eq:perp-time}
\eeq
where time $t=0$ when the caustic crosses the origin (i.e., the black hole position).

\subsection{Computing the Microlensed Spectrum}
\label{sec:numerics}

For an extended source such as a quasar accretion disk, we obtain the resulting microlensed flux $F$ by convolving the amplification map with the intensity map of the source. Using the fold caustic approximation of the map given by Equation~(\ref{eq:fold}), we get
\beq
F(E;\alpha_0,\beta_0)=\theta^{-2}\int\limits_{\rm disk} A[\,d_\perp(\alpha,\beta;\alpha_0,\beta_0)\,]\,I_{\rm obs}(E;\alpha,\beta) \,{\rm d}\alpha\,{\rm d}\beta\,,
\label{eq:flux}
\eeq
where the position of the caustic is defined by $(\alpha_0,\beta_0)$ and the factor $\theta$ converts plane-of-the-sky coordinates $(\alpha,\beta)$ to angular units.

For a caustic moving across the disk, the time dependence of the perpendicular coordinate $d_\perp$ is defined by Equation~(\ref{eq:perp-time}). For simplicity, we scale the observer-frame energy to the rest-frame energy of the iron line, defining $g_{\rm obs}\equiv E/E_{\rm Fe}$. If we now substitute for the observer-frame intensity from Equations~(\ref{eq:intensity-cont-obs}) and (\ref{eq:intensity-iron-obs}), we obtain the observed spectrum as a combination of the continuum flux
\beq
F^{\rm cont}(g_{\rm obs}E_{\rm Fe};t) = F_0^{\rm cont}\,g_{\rm obs}^{-\Gamma} \int\limits_{\rm disk} A[\,d_\perp(\alpha,\beta;t)\,]\, g^{\Gamma+2}(\alpha,\beta)\, r^{-q}(\alpha,\beta)\, {\rm d}\alpha\,{\rm d}\beta
\label{eq:flux-cont}
\eeq
and the flux from the iron line
\beq
F^{\rm Fe}(g_{\rm obs}E_{\rm Fe};t) = F_0^{\rm Fe}\int\limits_{\rm disk} A[\,d_\perp(\alpha,\beta;t)\,]\, g^3(\alpha,\beta)\,r^{-q}(\alpha,\beta)\, \delta\left[g_{\rm obs}-g(\alpha,\beta)\,\right]\, {\rm d}\alpha\,{\rm d}\beta\,.
\label{eq:flux-iron}
\eeq
The $F_0$ factors are combinations of constants, namely, $F_0^{\rm cont}=I_0^{\rm cont}\theta^{-2}E_{\rm Fe}^{-\Gamma}$ and $F_0^{\rm Fe}=I_0^{\rm Fe}\theta^{-2}E_{\rm Fe}^{-1}$. Note that unlike the $I_0$ values these constants have the same physical dimension. As seen from the dependence on $g_{\rm obs}$ in Equation~(\ref{eq:flux-cont}), the continuum part of the microlensed spectrum retains the same power-law shape, with amplitude modulated by the time-dependent integral. This is a consequence of the decoupled spatial and energy dependence of $I_{\rm obs}^{\rm cont}$ in Equation~(\ref{eq:intensity-cont-obs}).

On the other hand, the spatial and energy dependence of $I_{\rm obs}^{\rm Fe}$ in Equation~(\ref{eq:intensity-iron-obs}) are coupled through the argument of the delta function. In addition, the spatial coordinates are coupled with time in the argument of the amplification in Equation~(\ref{eq:flux-iron}). Hence, the iron line profile varies in amplitude as well as in shape during a microlensing event. Examining the delta function in Equation~(\ref{eq:flux-iron}), we see that the line profile extends from $g_{\rm min}$ to $g_{\rm max}$, where $g_{\rm min}$ and $g_{\rm max}$ are the minimum and maximum value, respectively, of the $g$-factor $g(\alpha,\beta)$ on the projected disk. For any value $g_{\rm obs}$ within this range, the delta function effectively reduces the two-dimensional integral to a line integral along the $g$-factor contour $g(\alpha,\beta)=g_{\rm obs}$.

It might seem straightforward to evaluate the integrals in Equations~(\ref{eq:flux-cont}) and (\ref{eq:flux-iron}) by inverse ray shooting, after replacing the continuous energy dependence by a sequence of $g_{\rm obs}$ intervals. However, this method is poorly suited for obtaining good spectral resolution and achieving sufficient accuracy for studying the evolution of fine spectral effects. Its drawbacks include the following: (1) integrating each ray as a geodesic in the Kerr metric back to the disk to obtain $r(\alpha,\beta)$ and $g(\alpha,\beta)$ is time consuming; (2) two-dimensional ray shooting is poorly suited for computing one-dimensional contour integrals; (3) the integrands in Equations~(\ref{eq:flux-cont}) and (\ref{eq:flux-iron}) are divergent along the caustic crossing the contours; (4) the small high-intensity region between the horizon and the $g$-factor peak (seen, e.g., in the right panel of Figure~\ref{fig:spec-int}) contributes to all energies within the spectral line, so that one would need a high density of rays to obtain reliable results. Overcoming these drawbacks would require high numbers of rays integrated back toward the disk, making ray shooting computationally cumbersome and numerically inefficient in this scenario.

Instead, we compute the microlensed spectrum given by Equations~(\ref{eq:flux-cont}) and (\ref{eq:flux-iron}) at a given time $t$ using an adaptive grid integration algorithm. We divide the projected disk into a grid of square cells and evaluate $r(\alpha,\beta)$ and $g(\alpha,\beta)$ for the corners of the squares by geodesic integration. For the continuum we bilinearly interpolate the mostly slowly varying term $g^{\Gamma+2}\,r^{-q}$ in each square. We then combine the result with the expression for the amplification and integrate over one coordinate analytically, followed by numerical integration over the second coordinate. Next, we refine the grid by a factor of two in either dimension and repeat the calculation in the same way. Finally, we estimate the absolute error for each of the original squares and the relative error with respect to the cumulative flux from all squares. If the estimated error in some square is larger than the threshold, we divide it and each of its eight immediate neighbors again into four smaller squares. In this way we preserve continuity of the interpolation over the full disk. We repeat this algorithm until all relative errors are lower than a required threshold or a maximum number of iterations is exceeded. Note that for the continuum it is sufficient to evaluate one integral for all photon energies $g_{\rm obs}$, due to the form of Equation~(\ref{eq:flux-cont}).

For the iron line we have to evaluate the integral in Equation~(\ref{eq:flux-iron}) separately for each chosen value of $g_{\rm obs}\in[g_{\rm min},g_{\rm max}]$. The procedure is similar to that used for the continuum, with a few alterations. In each square we bilinearly interpolate $g^3\,r^{-q}$ and $g_{\rm obs}-g$. We integrate only over the squares intersected by the interpolated $g$-factor contour $g(\alpha,\beta)=g_{\rm obs}$. Analytical integration over one coordinate is trivial, due to the presence of the delta function. Nevertheless, even in this case the integration over the second coordinate has to be performed numerically. For more details of the numerical procedure see~\cite{ledvina_heyrovsky15}.

In order to verify the reliability of our results, we check that the spectra converge to the unlensed line profile when the caustic is far from the disk center (in the limits $t\to\pm\infty$). This profile, illustrated by the blue curves in Figures~\ref{fig:spec-evol-0}--\ref{fig:spec-evol-infl}, can be also obtained by setting $A[\,d_\perp(\alpha,\beta;t)\,]=1$ in Equations~(\ref{eq:flux-cont}) and (\ref{eq:flux-iron}). Its shape is in agreement with the iron line profiles that can be found in the literature, such as in Figure 2 of \cite{laor91}, Figure 5 of \cite{fabian06}, or Figures 7--9 of \cite{dovciak_etal04}.

\section{VARIATIONS IN THE IRON K$\alpha$ LINE PROFILE}
\label{sec:spExample}

In this section we use the algorithm described in \S~\ref{sec:numerics} to compute sequences of spectra in the region of the iron K$\alpha$ line during a typical microlensing event. All examples presented here were calculated for caustic orientation $\psi=135^\circ$, strength $d_0=25\,r_{\rm g}$, direction of motion $\sigma_{\rm enter}=1$, and baseline amplification $A_0=1$, using Equations~(\ref{eq:perp-time}) and (\ref{eq:fold}). We kept the following accretion disk parameters fixed at sample values used for modeling accretion disk X-ray emission \citep{laor91,dovciak_etal04}: Kerr black hole spin $a=1$, spectral index~$\Gamma=2$, radial index~$q=3$, ratio between continuum and line radiation $F_0^{\rm cont}/F_0^{\rm Fe}=3$. In Boyer--Lindquist coordinates the disk extends from the ISCO at $r_{\rm in}=r_{\rm g}$ out to $r_{\rm out}=1000\,r_{\rm g}$. The radial integration limit $r_{\rm out}$ is positioned sufficiently far so that the emission is spatially constrained by the radial decline of the intensity rather than by a hard outer edge of the disk. Computations were performed for disk inclinations varying from $i=0^\circ$ to $i=85^\circ$, with most of the presented results corresponding to $i=70^\circ$. For the adaptive grid algorithm we used initial square side $10\,r_{\rm g}$, relative error threshold $10^{-4}$, and maximum number of iterations $18$.

\subsection{Spectral Evolution in the Course of a Microlensing Event}
\label{sec:evolution}

We first demonstrate the overall character of changes in the spectrum by computing the total disk-integrated flux $F=F^{\rm cont}+F^{\rm Fe}$ using Equations~(\ref{eq:flux-cont}) and (\ref{eq:flux-iron}) for a face-on and an inclined disk entering a caustic (Figure~\ref{fig:spec-evol-0} and Figure~\ref{fig:spec-evol-70}, respectively).

For either inclination the results are presented as a sequence of pairs of panels, showing the caustic position (left panel) and the corresponding spectrum (right panel). The caustic is plotted over a color map of the $g$-factor with the same contours as in Figure~\ref{fig:spec-int}. The caustic position is marked by the dark-orange line, and its inner side is indicated by the light-orange band. Hence, the emission from the part of the disk lying to the upper left from the caustic is amplified, with highest amplification along the caustic. The part lying to the lower right from the caustic is unamplified. More exactly, ``unamplified'' here and in the following text means amplified only by the constant baseline amplification $A_0$. Our specific choice $A_0=1$ is consistent with the literal meaning, yielding no excess flux when the caustic lies far from the disk center ($|t|\to\infty$).

The spectrum in each right panel is calculated for the corresponding time shown in the label in units of $t_{\rm g}$. The green curve represents the microlensed spectrum, plotted in terms of total flux in arbitrary units as a function of $g_{\rm obs}$, the photon energy relative to $E_{\rm Fe}$. For comparison we include the blue curve, which represents the spectrum of the same disk in absence of microlensing. The tick-mark spacing on the $g_{\rm obs}$ axis is equal to the $g$-factor contour spacing in the left panels, for easier interpretation of the iron line features.

\subsubsection{Face-on Disk ($\,i=0^\circ$)}
\label{sec:evol-0}

In Figure~\ref{fig:spec-evol-0} we present sample results for a face-on disk and three caustic positions. Since all emission from a face-on disk is redshifted ($g<1$), there is no contribution from the iron line in the spectra at $g_{\rm obs}>1$, where all flux comes from the continuum. Extending the power law from this region to lower energies shows the overall continuum spectrum, with all excess flux coming from the iron line. Note that due to the $g$-factor range (discussed in \S~\ref{sec:emission}), the full iron line extends from $g_{\rm obs}=0$ to $g_{\rm obs}=1$ for this inclination. The plotted spectra show only its more prominent high-energy part.

The top row illustrates the situation at time $t=-8.64\,t_{\rm g}$, before the caustic crosses the disk center. The continuum is modestly amplified, as indicated by the vertical offset of the green spectrum at energies $g_{\rm obs}>1$. The iron line profile is affected more dramatically, with a blunter edge at $g_{\rm obs}=1$ corresponding to emission far from the origin. In addition, there is a prominent new edge at $g_{\rm obs}=0.8$. This value corresponds to the lowest $g$-factor value in the region amplified by the caustic, as seen in the left panel. The part of the line at lower $g$-factor values, which originates from the innermost radii, remains unamplified.

The middle row corresponds to time $t=0\,t_{\rm g}$, at which the caustic crosses the disk center. The caustic amplifies emission exactly from half of the disk, down to the innermost $r=r_{\rm g}$. The amplification of the continuum is stronger because the caustic now covers part of the central region with strongest emission, as seen from Equation~(\ref{eq:intensity-cont-obs}). The iron line now has the blunt edge at $g_{\rm obs}=1$ seen before, but lacks additional features.

In the bottom row at time $t=8.64\,t_{\rm g}$ the caustic passed the center and lies at a position symmetrically opposite from the top row. The continuum is amplified even more than in the previous case because the central region is now fully inside and close to the caustic. The iron line exhibits the blunt edge at $g_{\rm obs}=1$, as well as a new sharp peak at $g_{\rm obs}=0.8$. This value now corresponds to the highest $g$-factor value with its contour fully inside the caustic, as seen in the left panel. For higher values, a part of the emitting region lies outside and thus remains unamplified.

At early times when the caustic is still far from the center ($t\to -\infty$), the amplification of the continuum is negligible since most of the disk lies outside the caustic, with $d_\perp<0$ in Equation~(\ref{eq:fold}). The line profile exhibits only a very weak additional edge at $g_{\rm obs}\to 1$. At late times when the caustic has moved far from the center ($t\to \infty$), the amplification of the continuum is negligible since the high-emission region lies far inside the caustic, with $d_\perp\gg d_0$ in Equation~(\ref{eq:fold}). The line profile exhibits a very weak additional peak at $g_{\rm obs}\to 1$. Hence, in both limits the spectrum converges to the blue non-microlensed spectrum.

Based on the above, we may describe the generic spectral evolution for a face-on disk entering the caustic. As the caustic approaches the disk, a new edge splits from the line edge at $g_{\rm obs}=1$ and shifts gradually to lower energies. Once the caustic reaches the ISCO, the edge disappears at zero energy. After the caustic crosses the central hole so that the ISCO is fully inside, a peak appears at zero energy. The peak gradually shifts to higher energies and eventually disappears at the $g_{\rm obs}=1$ line edge as the caustic moves far from the center.

\subsubsection{Inclined Disk ($\,i=70^\circ$)}
\label{sec:evol-70}

A similar sequence for an $i=70^\circ$ inclined disk entering a caustic is shown in Figure~\ref{fig:spec-evol-70}, with six positions proceeding from the top left to the bottom left pairs of panels and continuing down the right column. Clearly the line profile undergoes more complicated changes in this case. We first note that for such an inclination all $g$-factor contours pass through the high specific intensity area between the central black hole and the $g$-factor maximum lying to its left, as seen in the right panel of Figure~\ref{fig:spec-int}. Hence, most of the line radiation is emitted from this central part of the accretion disk rather than from parts farther away from the black hole as in the face-on disk case (left panel of Figure~\ref{fig:spec-int}). The strongest microlensing effect can thus be expected when the caustic crosses this region.

Since in this case there is a blue-shifted region in addition to the redshifted region (as indicated by the $g$-factor color), the limit in the spectrum beyond which there is no contribution from the iron line is shifted to $g_{\rm obs}=g_{\rm max}=1.304$, the maximum $g$-factor value for this inclination. The flux at $g_{\rm obs}>1.304$ comes purely from the amplified power-law continuum.

The top left panels correspond to the situation at time $t=-25\,t_{\rm g}$ before crossing the center, when the caustic is still far from the high-intensity region. The overall spectrum is only very weakly amplified. The only noticeable changes are in the iron line profile: there are two new edges, marked by the green arrows. The better visible blue-shifted edge at $g_{\rm obs}=1.15$ indicates the highest $g$-factor in the region amplified by the caustic. The barely noticeable redshifted edge at $g_{\rm obs}=0.98$ similarly indicates the lowest amplified $g$-factor. Parts of the line outside the interval delimited by these values are unamplified.

At time $t=-12.5\,t_{\rm g}$ in the middle left panels the caustic lies closer to the center; hence, the overall amplification of the continuum is more visible. The more prominent features visible in the line are analogous to those seen above. Here the part of the line $g_{\rm obs}\in(0.95,1.2)$ is amplified.

In the bottom left panels at $t=-3.86\,t_{\rm g}$ the caustic enters the central region. More specifically, it crosses the highest $g$-factor point lying within the innermost blue contour. This is clearly visible in the spectrum, with a very strong peak directly at the blue edge of the line at $g_{\rm obs}=1.304$. Within the limits of the plot, all of the line is amplified from this time onward. Besides the high overall amplification, no other spectral feature is visible in this plot. This special case of the caustic passing through the $g$-factor maximum is discussed further in \S~\ref{sec:specf-higher}.

The highest amplification of the six presented instants occurs in the top right panels at $t=-2.53\,t_{\rm g}$. The line profile exhibits a new prominent peak at $g_{\rm obs}=1.2$, similar to the peak in the bottom row of Figure~\ref{fig:spec-evol-0}. In this case contours with $g_{\rm obs}>1.2$ lie fully inside the caustic, while those with $g_{\rm obs}<1.2$ lie partly outside. In addition, there is a small peculiar bump near $g_{\rm obs}=1.02$, which we will return to briefly in \S~\ref{sec:specf-higher}.

At $t=-1.93\,t_{\rm g}$ the amplification remains high and the spectrum has two adjacent new peaks. The minor peak at $g_{\rm obs}=1.02$ has a similar nature to the peak visible at $t=-2.53\,t_{\rm g}$. Contours with a higher $g$-factor lie fully inside the caustic, those with a lower $g$-factor lie partly outside. However, the prominent peak at $g_{\rm obs}=0.9996$ is different. A closer inspection of the $g$-factor map reveals that contours on either side of the peak value intersect the caustic; no contour lies fully inside. Nevertheless, we find that the caustic is tangent to the $g=0.9996$ contour - but at its inflection point. We will discuss this special case in \S~\ref{sec:specf-higher} as well.

The bottom right panels at time $t=25\,t_{\rm g}$ correspond to the caustic positioned symmetrically opposite from the top left panels, after crossing the central region of the disk. Emission from all of the disk in the plot except the bottom right corner is amplified by the caustic. However, the high-intensity central region lies farther from the caustic. This implies a somewhat lower amplification, but also a drop in spatial resolution, since the variation of amplification across the central region is low. The effect can be seen in the spectrum: while the amplification is still significant, it is nearly uniform across the spectrum. The arrows mark the positions of tiny, hardly visible peaks.

At even later times the amplification drops, and the green spectrum merges with the blue spectrum. The same occurs at very early times, when the amplified part of the line shrinks to a small step around $g_{\rm obs}=1$.

\subsection{Generated Spectral Features: Peaks and Edges}
\label{sec:featuresPE}

The simulations in \S~\ref{sec:evol-0} and \S~\ref{sec:evol-70} show that the most typical features generated in the microlensed iron line profile are peaks and edges. Their appearance and location clearly depend on the geometry of the $g$-factor contours in the vicinity of the caustic. In order to explore this connection in more detail, we present in Figure~\ref{fig:spec-evol-peak} and Figure~\ref{fig:spec-evol-edge} zoomed-in details from Figure~\ref{fig:spec-evol-70} around a sample peak and edge, respectively. For simplicity, we plot only the line flux $F^{\rm Fe}$ in the right panels, ignoring the power-law continuum.

Peak evolution is illustrated in Figure~\ref{fig:spec-evol-peak}. The middle row corresponds to details from the $t=-2.53\,t_{\rm g}$ top right panels of the inclined-disk sequence in Figure~\ref{fig:spec-evol-70}. Shown in the right panel is the vicinity of the $g_{\rm obs}=1.2$ peak in the line profile up to the line edge at $g_{\rm obs}=1.304$. Enlarged in the left panel is the $g$-factor map region where the $g=1.2$ contour corresponding to the peak approaches the caustic. Dotted contours correspond to minor ticks on the $g_{\rm obs}$ axis in the right panel; solid contours correspond to major ticks. The top and bottom rows correspond to times preceding and succeeding the middle row, respectively.

Inspecting the situation at $t=-2.53\,t_{\rm g}$ in the middle row, we see that the caustic is exactly tangent to the $g=1.2$ contour, which lies on its inner amplified side. The position of the peak thus corresponds to the contour with the longest segment in the highest-amplification region along the caustic. Contours for $g$-factor values on either side of the peak have gradually shorter segments in this region.

The situation at the two bracketing times supports this interpretation. In the top row at $t=-2.63\,t_{\rm g}$ the caustic is tangent to the $g=1.223$ contour (not plotted), and the peak thus appears at this higher energy. In the bottom row at $t=-2.43\,t_{\rm g}$ the caustic is tangent to the $g=1.174$ contour (not plotted), and the peak appears at this lower energy. As the caustic sweeps across the disk in this sequence, the peak moves across the line profile, appearing at the energy corresponding to the $g$-factor contour instantaneously tangent to the caustic. This interpretation also implies that the speed and sense of the energy shift of the peak are given by the dot product of the $g$-factor gradient at the tangent point and the velocity of perpendicular caustic motion.

Edge evolution in Figure~\ref{fig:spec-evol-edge} is illustrated by the behavior around the $t=-12.5\,t_{\rm g}$ middle left panels of Figure~\ref{fig:spec-evol-70}. Shown in the middle right panel of Figure~\ref{fig:spec-evol-edge} is the vicinity of the $g_{\rm obs}=1.2$ edge in the line profile. The middle left panel shows the enlarged $g$-factor map region where the $g=1.2$ contour approaches the caustic. The top and bottom rows again correspond to times preceding and succeeding the middle row, respectively.

Concentrating first on the middle row at $t=-12.5\,t_{\rm g}$, we see the caustic is again exactly tangent to the $g=1.2$ contour. However, this time the contour lies on its outer, unamplified side. The position of the edge thus corresponds to the last contour reaching the caustic. Contours for $g$-factor values farther on the inner side have a small segment in the high-amplification region, while contours for values farther on the outer side have no segment in the amplified region.

In the top row at $t=-16.5\,t_{\rm g}$ the caustic is tangent to the $g=1.176$ contour (not plotted), and the edge thus appears at this lower energy. In the bottom row at $t=-8.5\,t_{\rm g}$ the caustic is tangent to the $g=1.238$ contour (not plotted), and the edge appears at this higher energy. In analogy to the case of the peak, as the caustic sweeps across the disk in this sequence, the edge moves across the line profile, appearing at the energy corresponding to the $g$-factor contour instantaneously tangent to the caustic. The speed and sense of the energy shift of the edge are computed in the same way as for the peak.

Note that in this sequence the edge is descending since the $g$-factor increases outward from the tangent point. If it decreased outward from the tangent point (e.g., if we flipped the orientation of the gradient), the line profile would have an ascending edge. This can be seen, for example, at $g_{\rm obs}=1.2$ in the top row of Figure~\ref{fig:spec-evol-0}.

From the peak and edge analysis above we infer that the position of these spectral features is directly given by the energy of the $g$-factor contour tangent to the caustic. A contour tangent from the inner side of the caustic generates a peak, while a contour tangent from the outer side generates an edge. Extrapolating further, we expect that the local profile of these features should be described by local properties of the $g$-factor map of the accretion disk at the tangent point. We test this hypothesis analytically in \S~\ref{sec:analytical}.

\subsection{Higher-order Spectral Features}
\label{sec:specf-higher}

Under special circumstances the tangent contour cannot be classified as reaching the caustic from the inner or outer side, leading to spectral features and their sequences other than the peaks and edges described in \S~\ref{sec:featuresPE}. One such case involves the point-like contour at the $g$-factor maximum, such as the one crossed by the caustic in the bottom left panels of Figure~\ref{fig:spec-evol-70}. Another case occurs when the tangent point is an inflection point of the contour, as shown in the middle right panels of Figure~\ref{fig:spec-evol-70}.

Spectral evolution during the crossing of the $g$-factor maximum is shown in Figure~\ref{fig:spec-evol-max}, detailing the situation around the $t=-3.86\,t_{\rm g}$ panels of Figure~\ref{fig:spec-evol-70}. In the top row at $t=-5.00\,t_{\rm g}$ the maximum lies outside the caustic, which is tangent from its outer side to the $g=1.291$ contour and thus forms an edge in the line profile. In the middle row at $t=-3.86\,t_{\rm g}$ the caustic directly crosses the maximum, forming a prominent asymmetric peak at the line edge. In the bottom row at $t=-3.00\,t_{\rm g}$ the maximum lies inside the caustic, which is now tangent from its inner side to the $g=1.275$ contour (not plotted) and thus forms a regular peak in the line profile.

In general, the crossing of a maximum of the $g$-factor is associated with a change from an edge to a peak (or vice versa) via a strong asymmetric peak. In the disk emission models used in this study the only local maximum (and the only local extremum) is the global maximum of the $g$-factor. Hence, the asymmetric peak occurs only at the high-energy edge of the line.

Figure~\ref{fig:spec-evol-infl} illustrates the spectral change during the tangent crossing of a $g$-factor contour inflection point, detailing the situation around the $t=-1.93\,t_{\rm g}$ panels of Figure~\ref{fig:spec-evol-70}. Note that the presented sequence occurs over a very short time interval of $0.08\,t_{\rm g}$, so that the shifting caustic position is barely noticeable in the left column. Each of the three line profiles includes a small peak at $g_{\rm obs}=1.020$ unrelated to the inflection crossing, from a contour tangent internally to the caustic beyond the area plotted in the left column. In the following we discuss the other line profile features.

In the top row at $t=-1.98\,t_{\rm g}$ the caustic is tangent from its inner side to the $g=1.0137$ contour (not plotted), forming a peak in the line profile. To the upper right of this tangent point the caustic is crossed from inside by the $g=1.0068$ contour (not plotted), which turns back to touch the caustic from outside to the lower left of the first contour's tangent point, forming an edge in the line profile. As the caustic advances to the inflection point at $t=-1.93\,t_{\rm g}$ in the middle row, the peak and edge merge to form a prominent asymmetric peak at $g_{\rm obs}=0.9996$, equal to the value of the $g$-factor contour tangent at its inflection point. In the bottom row at $t=-1.90\,t_{\rm g}$ no contour is tangent to the caustic in the vicinity of the crossed inflection point. As a result, the prominent peak disappears, leaving behind a bump at $g_{\rm obs}=0.9938$ corresponding to the contour with the longest segment in the high-magnification region along the caustic.

To summarize, the tangent crossing of a $g$-factor contour inflection point is associated with a change from an adjacent edge and peak pair via a strong asymmetric peak to a bump that eventually vanishes. Studying the full set of $g$-factor contours and plotting their inflection points yields curves running across the contours in the map (see further in \S~\ref{sec:peakmaps}). The caustic may cross any of these points at different energies. However, we stress that the described sequence with the strong peak occurs only if the caustic is tangent to the local $g$-factor contour passing through the point.

Finally, we note that the small bump visible at $g_{\rm obs}=1.02$ in the top right panels of Figure~\ref{fig:spec-evol-70} has a similar origin to the bump in the bottom row of Figure~\ref{fig:spec-evol-infl}. It arises outside the plotted area from a nontangent contour closely approaching the high-amplification region along the caustic.

\section{ANALYTICAL MODEL OF GENERATED FEATURES}
\label{sec:analytical}

The results presented in section \textsection~\ref{sec:spExample} indicate that the prominent spectral features generated by microlensing have a local origin, since they arise from points on the disk where the caustic is tangent to $g$-factor contours. Here we explore these features analytically by studying the contribution to the line flux from a small circular area around such a point. As we demonstrate below, such a model is sufficient for modeling the features described in \S~\ref{sec:featuresPE}.

We take a small circular source with radius $\rho$ centered on a point $(\alpha_0,\beta_0)$ lying on the $g$-factor contour $g=g_0$. In order to perform the subsequent integrations, it is useful to transform to local coordinates $(\alpha,\beta) \rightarrow (x,y)$ by shifting the origin to the center of the circle and rotating the axes by an angle $\xi$ so that $x$ lies along the $g_0$ contour. For a caustic with a general orientation positioned so that it passes through the origin, the perpendicular coordinate given by Equation~(\ref{eq:perp-origin}) expressed in terms of $x,y$ is
\beq
d_\perp(x,y)=x\,\cos(\psi-\xi)+y\,\sin(\psi-\xi)\,.
\label{eq:perp-circle}
\eeq

Since we are interested in the leading-order result in terms of spectral separation from the feature, we compute for this ``circular source'' only the excess flux $\Delta F^{\rm Fe}$ due to the divergent part of the amplification in Equation~(\ref{eq:fold}). The line flux from Equation~(\ref{eq:flux-iron}) thus can be written as
\beq
F^{\rm Fe}=F_{\rm base}^{\rm Fe}+\Delta F^{\rm Fe}\,,
\label{eq:flux-split}
\eeq
where $F_{\rm base}^{\rm Fe}$ includes the flux from the rest of the disk plus the $A_0$ --- amplified flux from the circle. The excess flux can be computed from
\beq
\Delta F^{\rm Fe}(g_{\rm obs}E_{\rm Fe})= F_0^{\rm Fe}\,\sqrt{d_0}\int\limits_{\rm circle} \frac{g^3(x,y)\,r^{-q}(x,y)}{\sqrt{d_\perp(x,y)}}\,H[d_\perp(x,y)]\,\delta\left[g_{\rm obs}-g(x,y)\,\right]\, {\rm d}x\,{\rm d}y\,.
\label{eq:flux-excess}
\eeq

In order to simplify the argument of the delta function, we replace the $g$-factor by its expansion around the origin. For purposes of illustration it is sufficient to use the simplest expansion with curved contours,
\beq
g(x,y) \approx g_0 + g_y\,y + \frac12\, g_{xx}\,x^2\,,
\label{eq:gfapprox}
\eeq
where $g_y$ and $g_{xx}$ are the respective partial derivatives of the $g$-factor at the origin\footnote{It can be shown that adding the $g_{yy}$ and $g_{xy}$ terms of a general Taylor expansion does not affect the leading-order features derived further, provided that $g_{y}$ and $g_{xx}$ are nonzero.}. Note that if these are nonzero, the contours $g(x,y)={\rm const}$ of this model are parabolas shifted along their common symmetry axis $y$.

We can now utilize the delta function to perform the integration along $y$ in Equation~(\ref{eq:flux-excess}), yielding
\beq
\Delta F^{\rm Fe}(g_{\rm obs}E_{\rm Fe})= F_0^{\rm Fe}\,\sqrt{d_0}\;\frac{g_{\rm obs}^3}{|g_y|}\int\limits_{\rm x-range} \frac{r^{-q}(x,y(x))}{\sqrt{d_\perp(x,y(x))}} \,H[d_\perp(x,y(x))]\,{\rm d}x\,,
\label{eq:flux-excess-1}
\eeq
where the remaining integration is to be performed over the $x$-range of the $g=g_{\rm obs}$ contour on the circle. The delta function replaced the $g$-factor by $g_{\rm obs}$, generated the derivative in the denominator, and replaced $y$ by its value on the $g(x,y)=g_{\rm obs}$ contour:
\beq
y(x)=-g_y^{-1}\left(\frac12\, g_{xx}\,x^2+g_0-g_{\rm obs}\right)\,.
\label{eq:contour-y}
\eeq
Of the three remaining functions in the integrand, the disk coordinate will vary much more slowly than the divergent denominator or the Heaviside function. For computing the leading-order term we can safely replace it by its value $r_{\rm circ}$ at the origin of the circle, leading to
\beq
\Delta F^{\rm Fe}(g_{\rm obs}E_{\rm Fe})= F_0^{\rm Fe}\,r_{\rm circ}^{-q}\,\sqrt{d_0}\;\frac{g_{\rm obs}^3}{|g_y|}\int\limits_{\rm x-range} \frac{H[d_\perp(x,y(x))]}{\sqrt{d_\perp(x,y(x))}} \,{\rm d}x\,.
\label{eq:flux-excess-2}
\eeq

In the following two subsections we compute the excess flux for caustic orientation tangent to the $g_0$ contour at the origin. The caustic then coincides with the $x$-axis, and if we set $\psi=\xi-\pi/2\,$, the inner side corresponds to $y<0$ and Equation~(\ref{eq:perp-circle}) reduces to $d_\perp(x,y)=-y$. If we then use Equation~(\ref{eq:contour-y}), the excess flux for the tangent caustic simplifies to
\beq
\Delta F^{\rm Fe}(g_{\rm obs}E_{\rm Fe})= F_0^{\rm Fe}\,r_{\rm circ}^{-q}\,\sqrt{d_0}\;\frac{g_{\rm obs}^3}{|g_y|}\int\limits_{\rm x-range} \sqrt{\frac{g_y}{\frac12\, g_{xx}\,x^2+g_0-g_{\rm obs}}} \,{\rm d}x\,,
\label{eq:flux-excess-tangent}
\eeq
where the integration range is further limited by the requirement of a positive expression under the square root. In \S~\ref{sec:mod-peaks} we study the case $g_{xx}/g_y>0$, when the $g_0$ contour is tangent from inside; in \S~\ref{sec:mod-edges} the case $g_{xx}/g_y<0$, when the $g_0$ contour is tangent from outside the caustic.

In \S~\ref{sec:mod-general_angle} we compute the excess flux for an arbitrary caustic orientation $\psi$, in order to demonstrate the correspondence between the location of the studied features in the line profile and contours tangent to the caustic.

The final two subsections are devoted to the spectral peaks and edges generated by microlensing. In \S~\ref{sec:peakmaps} we map the strength of the generated features for a caustic passing through a given point on the disk. \S~\ref{sec:accuracy} explores the range of applicability of the simplest peak model.

\subsection{Peaks}
\label{sec:mod-peaks}

A contour tangent from inside the caustic (i.e., from negative $y$ in our example) requires $g_{xx}/g_y>0$, as can be seen from Equation~(\ref{eq:gfapprox}). This configuration is illustrated in the top left panel of Figure~\ref{fig:circ-model-size}, in which the parabolic contours are plotted over the source divided into concentric rings, with the outer radius normalized to unity. The $g(x,y)=g_0$ contour passes through the center.

The remaining simple integration yields an inverse hyperbolic function with the type depending on the sign of $(g_0-g_{\rm obs})/g_y$: inverse hyperbolic sine for positive, inverse hyperbolic cosine for negative. The integration limits for contours near $g_0$ depend on the position of the contour, as seen on the sketch in Figure~\ref{fig:circ-model-size}. The contours farther inside the caustic extend from edge to edge of the source, while those protruding outside the caustic have to be integrated in two segments extending from the edge to the caustic, omitting the central part. The $x$ coordinate of the intersection of these contours with the caustic can be obtained by setting the denominator in the integrand of Equation~(\ref{eq:flux-excess-tangent}) equal to zero. The intersection with the edge of the source can be obtained by substituting from Equation~(\ref{eq:contour-y}) in $x^2+y^2=\rho^2$.

Rather than providing all the bulky but straightforward intermediate expressions explicitly, we present here the leading-order term when $g_{\rm obs} \rightarrow g_0$, i.e., the line profile in the vicinity of $g_0$. Interestingly, this term has the same form for all the specific cases:
\beq
\Delta F^{\rm Fe}(g_{\rm obs}E_{\rm Fe})\approx - I\,\sqrt{\frac{2\,d_0}{g_y\,g_{xx}}}\,\ln|g_{\rm obs}-g_0|= -P \sqrt{d_0}\,\ln|g_{\rm obs}-g_0|\,,
\label{eq:peak-approx}
\eeq
where the constant $I=F_0^{\rm Fe}\,r_{\rm circ}^{-q}\,g_0^3$ is independent of the contour geometry and of the caustic properties. We may separate the caustic strength parameter from the contour geometry and define the peak strength
\beq
P=I\sqrt{2/|g_y\,g_{xx}|}\,,
\label{eq:peak-strng}
\eeq
which depends purely on the local disk properties, including the local shape of the $g$-factor contour.

The result in Equation~(\ref{eq:peak-approx}) shows the logarithmic character of the peak generated at $g_0$, the $g$-factor of the contour tangent to the caustic from inside. This result corroborates the behavior seen in the numerical simulations in \S~\ref{sec:featuresPE} and Figure~\ref{fig:spec-evol-peak}.

In the top row of Figure~\ref{fig:circ-model-size} the spectra in the two right panels show the numerically integrated excess flux generated by the caustic. In this example $g_y=-0.1$ (the $g$-factor decreases upward in the $y$-direction) and $g_{xx}=-0.3$. In the middle panel the color-coded widening profiles from blue to yellow correspond to increasing the source size from the small inner circle to the full outer circle. In the right panel the color-coded increasing profiles from yellow to blue correspond to an annular source starting from the outer ring, with the inner radius gradually shrinking to zero. Both plots clearly show that the region of the source directly at the tangent point is responsible for the (logarithmic) divergence at $g_0$.

If we swap signs of both relevant $g$-factor derivatives, the shape of the contours in the left panel remains the same; only the $g$-factor increases upward in the $y$-direction, and the spectra are mirrored vertically around~$g_0$.

\subsection{Edges}
\label{sec:mod-edges}

The $g_0$ contour is tangent from outside the caustic (i.e., from positive $y$ in our example) if the derivatives fulfill the condition $g_{xx}/g_y<0$. This configuration is illustrated in the bottom left panel of Figure~\ref{fig:circ-model-size}, with the $g(x,y)=g_0$ contour passing through the center.

The remaining simple integration in Equation~(\ref{eq:flux-excess-tangent}) depends on the sign of $(g_0-g_{\rm obs})/g_y$: we get inverse hyperbolic sine for positive, zero for negative. In the former case the integration limits for contours near $g_0$ extend from the caustic to the caustic, as seen on the sketch in Figure~\ref{fig:circ-model-size}. In the latter case the contours lie entirely outside the caustic, hence the zero contribution.

For both cases we can write the leading-order term when $g_{\rm obs} \rightarrow g_0$ in the form
\beq
\Delta F^{\rm Fe}(g_{\rm obs}E_{\rm Fe})\approx I\,\sqrt{\frac{2\,d_0}{-g_y\,g_{xx}}}\,\pi\,H\left[\frac{g_0-g_{\rm obs}}{g_y}\right]= P \sqrt{d_0}\,\pi\,H\left[\frac{g_0-g_{\rm obs}}{g_y}\right]\,,
\label{eq:edge-approx}
\eeq
where $H$ is the Heaviside step function, and the constant $I$ has the same definition as in \S~\ref{sec:mod-peaks}. By separating the caustic strength parameter from the contour geometry, we get the same parameter $P$ defined in Equation~(\ref{eq:peak-strng}), depending purely on the local disk properties, including the local shape of the $g$-factor contour. Instead of peak strength, here it appears as a measure of the step size.

The result in Equation~(\ref{eq:edge-approx}) shows the step-like character of the edge generated at $g_0$, the $g$-factor of the contour tangent to the caustic from outside. This result corroborates the behavior seen in the numerical simulations in \S~\ref{sec:featuresPE} and Figure~\ref{fig:spec-evol-edge}.

In the bottom row of Figure~\ref{fig:circ-model-size} the spectra in the two right panels show the numerically integrated excess flux generated by the caustic. In this example $g_y=0.1$ (the $g$-factor increases upward in the $y$-direction) and $g_{xx}=-0.3$. In the middle panel the color-coded widening profiles from blue to yellow correspond to increasing the source size from the small inner circle to the full outer circle. In the right panel the color-coded increasing profiles from yellow to blue correspond to an annular source starting from the outer ring with the inner radius gradually shrinking to zero. Both plots show that the region of the source directly at the tangent point is responsible for the step-like high-energy edge at $g_0$.

Just like in \S~\ref{sec:mod-peaks}, if we swap signs of both relevant $g$-factor derivatives, the shape of the contours in the left panel remains the same; only the $g$-factor decreases upward in the $y$-direction, and the spectra are mirrored vertically around~$g_0$ with the step forming a low-energy edge.

\subsection{Changing Features with Caustic Orientation}
\label{sec:mod-general_angle}

For a caustic crossing the contour at the origin under a general angle, we substitute the general expressions from Equations~(\ref{eq:perp-circle}) and (\ref{eq:contour-y}) in Equation~(\ref{eq:flux-excess-2}) and obtain the excess flux in the form
\beq
\Delta F^{\rm Fe}(g_{\rm obs}E_{\rm Fe})= F_0^{\rm Fe}\,r_{\rm circ}^{-q}\,\sqrt{d_0}\;\frac{g_{\rm obs}^3}{|g_y|}\int\limits_{\rm x-range} \sqrt{\frac{g_y}{g_y\,x\,\cos(\psi-\xi)-(\frac12\, g_{xx}\,x^2+g_0-g_{\rm obs})\,\sin(\psi-\xi)}} \;{\rm d}x\,.
\label{eq:flux-excess-general}
\eeq
where the integration range is limited to the positions on the circular source with a positive expression under the square root. The expression in the denominator of the integrand can be rearranged by collecting the terms depending on $x$ to form a complete square, yielding
\beq
\Delta F^{\rm Fe}(g_{\rm obs}E_{\rm Fe})= F_0^{\rm Fe}\,r_{\rm circ}^{-q}\,\sqrt{d_0}\;\frac{g_{\rm obs}^3}{|g_y|}\int\limits_{\rm x-range} \sqrt{\frac{-g_y\,/\sin(\psi-\xi)}{\frac12\, g_{xx}\,[x-x_{\rm tp}(\psi)]^2+g_{\rm tp}(\psi)\,-g_{\rm obs}}} \;{\rm d}x\,,
\label{eq:flux-excess-general-1}
\eeq
where
\beq
g_{\rm tp}(\psi)=g_0-\frac{g_y^2}{2\,g_{xx}}\cot^2(\psi-\xi)\,,\;\;{\rm and}\;\;x_{\rm tp}(\psi)=\frac{g_y}{g_{xx}}\cot(\psi-\xi)
\label{eq:tangentpoint}
\eeq
are the $g$-factor of the contour that is tangent to the caustic and the position of the tangent point, respectively.

Comparison with Equation~(\ref{eq:flux-excess-tangent}) shows that the resulting profiles will be very similar. However, the peak or edge will appear at $g_{\rm tp}$ instead of $g_0$, as long as the tangent point $x_{\rm tp}$ lies within the circular source. The type of feature is now given by the sign of $g_{xx}\,\sin(\psi-\xi)/g_y$.

A peak occurs if the contour is tangent from inside the caustic, which requires $g_{xx}\,\sin(\psi-\xi)/g_y<0$. The leading-order term when $g_{\rm obs} \rightarrow g_{\rm tp}$ gives a flux excess
\beq
\Delta F^{\rm Fe}(g_{\rm obs}E_{\rm Fe})\approx - I\,\sqrt{\frac{-2\,d_0}{g_y\,g_{xx}\,\sin(\psi-\xi)}}\,\ln|g_{\rm obs}-g_{\rm tp}|\,,
\label{eq:peak-approx-general}
\eeq
where the constant $I$ has the same form as in \S~\ref{sec:mod-peaks}. Changing the caustic orientation away from the tangent $\psi=\xi-\pi/2$ shifts the peak position $g_{\rm tp}$ and boosts the strength of the peak. This increase can be explained by the contour staying longer near the caustic.

An edge occurs if the contour is tangent from outside the caustic, with $g_{xx}\,\sin(\psi-\xi)/g_y>0$. The leading-order term when $g_{\rm obs} \rightarrow g_{\rm tp}$ gives a flux excess
\beq
\Delta F^{\rm Fe}(g_{\rm obs}E_{\rm Fe})\approx I\,\sqrt{\frac{2\,d_0}{g_y\,g_{xx}\,\sin(\psi-\xi)}}\,\pi\,H\left[\frac{g_{\rm obs}-g_{\rm tp}}{g_y}\,\sin(\psi-\xi)\right]\,.
\label{eq:edge-approx-general}
\eeq
The dependence on caustic orientation is similar here, with the amplitude of the step increasing when rotating the caustic away from the tangent $\psi=\xi-\pi/2$.

A sequence of numerically integrated flux excess profiles for different caustic orientations is shown in Figure~\ref{fig:circ-model}, rotating gradually from $\psi=\xi-\pi/2$ in the first column to $\psi=\xi+\pi/2$ in the last column. The orientation in the first column produces a peak at $g_0$, as seen in the top row of Figure~\ref{fig:circ-model-size}. In the second column the internally tangent contour is marked by the dot-dashed line in the upper sketch. In this case $g_{\rm tp}>g_0$, as can be seen from the peak position in the bottom panel. In the third column no contour is tangent to the caustic within the circular source, and thus the corresponding profile has no caustic-related feature. In the fourth column an externally tangent contour is marked by the dot-dashed line in the top panel. The corresponding edge can be seen at the position $g_{\rm tp}>g_0$ in the bottom panel. In this case the edge looks like a peak from its red side. However, this is merely an artifact of the cutoff at the edge of the circle, since contours with lower $g_{\rm obs}$ would approach the caustic outside the circular area. Finally, the last column shows the edge at $g_0$ demonstrated in detail in the bottom row of Figure~\ref{fig:circ-model-size}.

We can now safely conclude that the peaks or edges generated in the line profile by the microlensing caustic directly correspond to $g$-factor contours tangent to the caustic from its inner or outer side, respectively.

\subsection{Mapping the Strength of Generated Features}
\label{sec:peakmaps}

The peak strength $P$ defined by Equation~(\ref{eq:peak-strng}) is obtained by multiplying two factors. The intensity factor $I$, which depends on the assumed disk emissivity law, is equal up to a conversion constant to the color maps shown in Figure~\ref{fig:spec-int} for two inclinations. The square-root factor in Equation~(\ref{eq:peak-strng}) depends purely on the local shape of the $g$-factor contours, which is in turn given by the geometric setup of a Keplerian disk orbiting in the equatorial plane of a Kerr black hole.

In Figure~\ref{fig:peak-strg} we present maps of the peak strength $P$ for six disk inclinations $i$ ranging from the face-on $0^\circ$ to a nearly edge-on $85^\circ$ orientation. We construct the maps by taking each point on the projected disk, extending the caustic through it as a  tangent of the local $g$-factor contour, computing the local parabolic approximation of the contour following Equation~(\ref{eq:gfapprox}) with the caustic defining the $x$-axis, and using the obtained $g$-factor derivatives to compute the peak strength $P$ from Equation~(\ref{eq:peak-strng}). As discussed in the previous sections, the caustic generates a peak or edge in the line profile when the contour lies on its amplified or unamplified side, respectively. The computed $P$ values correspond to either feature. The obtained maps can tell us which regions of the projected disk may produce prominent peaks or edges when crossed by a favorably oriented caustic, and which regions may yield only weak features.

For the face-on disk the map is axially symmetric. With increasing inclination we can see changes in the geometry of the $g$-factor contours. Up to inclination $10^\circ$ there is no prominent change in peak strengths (the $g$-factor maximum lies far from the black hole). Increasing the inclination further up to $30^\circ$ leads to some $g$-factor contours locally changing from convex to concave. This generates a curve in the map with divergent peak strength, since $g_{xx}$ vanishes at $g$-factor contour inflection points (see description in \S~\ref{sec:specf-higher}).

The next panel with $70^\circ$ inclination already shows the $g$-factor maximum inside the plotted area. At this point the peak strength diverges as well, here due to the vanishing of $g_y$ (see description in \S~\ref{sec:specf-higher}). Note also the increased peak strength around coordinates $(15\,r_{\rm g}, 10\,r_{\rm g})$. In this area the contours are very flat, and a further increase to $i=75^\circ$ generates here two new inflection points on each affected contour. Thus, at inclination $75^\circ$ we can see three different regions where peak strength $P$ diverges: two inflection curves (curves connecting the inflection points of $g$-factor contours) and the $g$-factor maximum. The last panel with inclination $85^\circ$ displays a third inflection curve formed above the horizontal axis extending from the center to the left.

The obtained plots clearly show that for higher inclinations one may expect stronger variations in generated peaks or edges as the caustic progresses across the disk. Note, however, that the requirement of tangency implies that for higher inclinations strong generated features occur for specific caustic orientations.

Finally, we recall that the peak strength is modulated by the intensity factor $I$, which affects the amplitude but not the occurrence or spectral position of the peak. A different emissivity law would change the color maps but not the contours shown in Figure~\ref{fig:spec-int}. This would lead primarily to a relative boost or suppression of peak strength in the central vs. outer regions of the disk.

\subsection{Accuracy of peak model}
\label{sec:accuracy}

In \S~\ref{sec:mod-peaks} we derived an analytical expression for the line profile in the vicinity of the microlensing-generated peak. Here we study its applicability by evaluating how accurately it fits the line flux in a microlensed spectrum computed following Equation~(\ref{eq:flux-iron}), as described in \S~\ref{sec:numerics}.

We perform tests for caustic tangent points taken from a regular $512\times512$ grid spanning an $\alpha,\beta\in (-30\,r_{{\rm g}},\,30\,r_{{\rm g}})$ square in the plane of the sky. For each point we calculate first and second derivatives of the $g$-factor with respect to $\alpha,\beta$. Using the notation introduced in Equation~(\ref{eq:perp-origin}), these define the orientation of the caustic tangent to the $g$-factor contour $g(\alpha,\beta)=g_0$ passing through the point,
\beq
(\cos{\psi},\sin{\psi})=\pm(g_{\alpha},g_{\beta})\,/\,[\,g_{\alpha}^2+g_{\beta}^2\,]^{1/2}\,,
\label{eq:orientation}
\eeq
where the sign is chosen so that the angle $\psi$ defining the inner side of the caustic points toward the center of curvature of the contour (identified using the second derivatives). We then compute the microlensed line flux $F^{\rm Fe}(g)$ from Equation~(\ref{eq:flux-iron}) for a disk with inclination $i=70^\circ$ and a caustic with $d_0=25\,r_{\rm g}$ and $A_0=1$, using Equation~(\ref{eq:perp-origin}) for the distance $d_\perp$.

The analytical model we use to fit the peak region,
\beq
F_{\rm m}^{\rm Fe}(g) = - C_1\,\ln|g-g_{\rm m0}| + C_2 + C_3\,(g-g_{\rm m0})\,
\label{eq:peak-model}
\eeq
is based on the result from Equation~(\ref{eq:peak-approx}) superimposed on a linear background. We obtain the model parameters $C_1,C_2,C_3,g_{\rm m0}$ by a least-squares fit to the line flux $F^{\rm Fe}(g)$ performed over the spectral interval $[g_0-\Delta g, g_0+\Delta g]$. We measure the goodness of fit as a function of increasing half-width $\Delta g$ by the relative approximation error
\beq
\epsilon\,(\Delta g) = \sqrt{\frac{\int[F_{\rm m}^{\rm Fe}(g)-F^{\rm Fe}(g)]^2\,{\rm d} g} {\int F^{\rm Fe}(g)^2\, {\rm d} g}}\,,
\label{eq:error}
\eeq
where we integrate from $g_0-\Delta g$ to $g_0+\Delta g$. We present the obtained results as a function of tangent point position in the form of two plots in Figure~\ref{fig:peak-accuracy}.

In the left panel of Figure~\ref{fig:peak-accuracy} we present a color map of the half-width of the fitted interval at which the error $\epsilon\,(\Delta g)$ reaches $1\%$. For better orientation we plot the contour $\Delta g=0.05$. Clearly, for tangent points in most of the plot the spectrum can be fitted over an even wider interval with $1\%$ accuracy. Note that there are regions where the map changes abruptly. This is caused by the dependence $\epsilon\,(\Delta g)$ not being monotonic. In one pixel there might be a local maximum just above $1\%$ and in a neighboring one just under $1\%$, causing a jump in the width of the fitted interval.

In the complementary right panel we present a color map of the relative error $\epsilon\,(0.05)$ obtained for the fixed half-width $\Delta g = 0.05$. Here we plot the contour $\epsilon=1\%$, which should closely correspond to the contour in the left panel. The small differences are caused by the nonmonotonic dependence $\epsilon\,(\Delta g)$ mentioned above. The plot shows that for most tangent points in the studied area the spectrum can be fitted over the interval $[g_0-0.05, g_0+0.05]$ with better than $1\%$ accuracy. For the iron K$\alpha$ line this interval corresponds to $\pm 0.3\,\rm keV$ around the microlensing-generated peak.

In both panels of Figure~\ref{fig:peak-accuracy} we see several areas inside the plotted contours where the quality of the fit decreases. Close to the horizon the $g$-factor contours change rapidly within a small area; hence, our approximation has limited validity. The area at upper right is caused by very flat $g$-factor contours, not adequately approximated by parabolas. The band running from $\alpha = -10\,r_{\mathrm g}$ at the lower edge to $\alpha = -10\,r_{\mathrm g}$ at the upper edge is caused by the presence of two very close peaks, which cross each other in the middle of the band. An outer band running from $\alpha = -25\,r_{\mathrm g}$ at the lower edge to $\alpha = -18\,r_{\mathrm g}$ at the upper edge surrounds the inflection curve, at which the contours are not parabolic. Finally, the area to the left of the center around the maximum of the $g$-factor has elliptic contours, with parabolic approximation possible only in very small sections.

We conclude that with the exception of these areas, microlensing-generated peaks from most parts of the central accretion disk are described by the model in Equation~(\ref{eq:peak-model}) with an accuracy better than $1\%$ within a $g$-factor interval of at least $\pm 0.05$ around the peak.

\section{DISCUSSION}
\label{sec:discussion}

In the presented results we discussed in detail locally generated spectral features, arising at a point where the caustic is tangent to a $g$-factor contour. For an arbitrarily positioned and oriented caustic there may be one such point (top and bottom panels of Figure~\ref{fig:spec-evol-0}), several tangent points on different contours (most of the panels of Figure~\ref{fig:spec-evol-70}), several tangent points on the same contour (e.g., for the disk in Figure~\ref{fig:spec-evol-70} a near-vertical caustic just to the left of the horizon, tangent to the weakly redshifted contours), or exceptionally no tangent point (middle panel of Figure~\ref{fig:spec-evol-0}). The line profile in the case of several points on different contours just exhibits several features at different energies, as discussed in \S~\ref{sec:evol-70}. Several points on the same contour would produce a superposition of several features at the same energy: a logarithmic peak or a step function with strength given by adding the individual strengths from the different points. As the caustic crosses this special alignment, the individual features would move apart in the profile. In the unlikely case of an internal and external tangent point on the same contour, the line profile formed by superposition of a peak and a step at the same energy would resemble the inflection-crossing profile shown in the middle panel of Figure~\ref{fig:spec-evol-infl}.

More generally, we concentrated on explaining the line profile for a given static position of the caustic. While we illustrated sample sequences of changes during a caustic-crossing microlensing event, we did not study the frequency of occurrence of different features or their combinations for arbitrary caustic orientations. From the $g$-factor contour geometries in Figure~\ref{fig:peak-strg} it is clear that peaks and edges will occur in any caustic crossing. Any caustic would also cross the $g$-factor maximum, generating temporarily the strong peak at the high-energy edge, as seen in Figure~\ref{fig:spec-evol-max}. However, the maximum peak strengths seen in Figure~\ref{fig:peak-strg}, which are generated in tangent crossings at $g$-factor contour inflection points, will not occur in all crossings. For example, in the $i=70^\circ$ case, more horizontally oriented caustics would cross all inflection points at a nontangent angle. An analysis of the frequency of occurrence of different combinations of features is beyond the scope of this work.

The model used to derive the analytical shapes of the features has its limitations. For example, we assumed constant specific intensity close to the tangent point. This assumption could fail if the specific intensity exhibited abrupt changes near the point without similar effects on the $g$-factor contours at a correspondingly small spatial scale. In principle, one could derive the same analytical shapes using an even smaller circular area around such a point. Nevertheless, a strong localized intensity variation would clearly introduce changes at a finer scale than the overall behavior dominated by the geometry of the $g$-factor contours.

The analytical models derived for the peak and edge fail if $g_y=0$ (e.g., near the $g$-factor maximum) or $g_{xx}=0$ (e.g., near $g$-factor inflection points). For these cases it is necessary to include more terms in the local $g$-factor expansion in Equation~(\ref{eq:gfapprox}), and the more complicated calculations do not yield simple analytical results. Higher-order terms should be included also in cases when the $g$-factor exhibits more complex changes in the vicinity of the tangent point. This can be seen, for example, in the high-inclination $i=85^\circ$ case in Figure~\ref{fig:peak-strg} for points close to the horizontal axis and close to the horizon. Nevertheless, all these are special cases not affecting the overall behavior.

The results presented here were computed for an accretion disk around a maximally rotating Kerr black hole with spin parameter $a=1$. Similar computations can be performed for a black hole with any other value of $a$, including the $a=0$ Schwarzschild black hole. In these cases, however, we may expect one systematic difference. Unlike in the $a=1$ case, the ISCO (at which we assume the inner radius of the disk) does not coincide with the horizon. As a result, we may expect additional features depending on the position of the caustic and the $g$-factor contour tangent point relative to the ISCO. We will explore these effects in a further study.

In our disk emission model we assumed simple laws for the rest-frame continuum given by Equation~(\ref{eq:intensity-cont-em}) and for the line by Equation~(\ref{eq:intensity-iron-em}). While we demonstrated results for fiducial values of the spectral index $\Gamma$ and radial index $q$, it is simple to produce results for any other assumed values. However, one may assume other emission laws, include directional dependence of rest-frame emission \citep[a.k.a. ``limb darkening"; see][]{beckwith_done04,fukue_akizuki06}, consider different configurations of the emission region than our assumed thin disk \citep[e.g.,][]{chen_etal13a}, or include higher-order images of the disk formed by the black hole near the horizon \citep{beckwith_done05}. Nevertheless, most of these changes would just lead to different geometries of the $g$-factor contours. As described in this work, due to the geometric origin of the generated spectral features, a microlensing caustic would again produce a sequence of peaks and edges, depending on the location and sequence of tangent contours. The differences in the specific intensity of the models would mostly lead to changes in the relative strengths of these features. As mentioned above, additional effects could be expected in the case of spatially abrupt variations in rest-frame intensity (e.g., due to spots or sharp edges). But only models without an overall continuous pattern of $g$-factor contours (e.g., with a more stochastic nature of the emitting region) would lead to systematically different behavior.

As a final remark on the emission side of our simulations, we presented calculations only for the Fe K$\alpha$ line, which is often the most prominent line in the X-ray spectra of quasars. However, an identical analysis can be performed for any other emission line that has a local rest-frame emission width negligible in comparison with the relativistic broadening mechanisms affecting the observed line profile.

Turning to our microlensing caustic model, a linear fold caustic is a reasonable first approximation for a compact source such as the X-ray-emitting inner region of the quasar. In some cases a linear caustic seems to be a good model even for the optically emitting region, which is larger by an order of magnitude \citep{mediavilla_etal15}. The next improvement of the model would include curvature of the caustic. For most situations we do not expect significant changes in the local behavior, since the amplification in the tangent case would locally depend on the distance between a parabolic contour and a parabolic caustic. In most cases the result would correspond to a straightened version of the caustic, and the strength of the generated peak or edge would depend on the relative curvature of the contours from the caustic. Clearly, if the curvatures are nearly equal at the tangent point, the generated peak or edge may be particularly strong.

The situation will be very different if the quasar is crossed by a cusp of the caustic. The cusp itself provides stronger spatial resolution, and the strength of the adjacent folds gradually increases toward it. Such a crossing is computationally more demanding but numerically straightforward, although without much prospect for deriving analytical properties of generated features. The same holds for the limit of the general microlensing caustic network, relevant for emitting sources with angular size larger than the structures of the network. However, proceeding further into this regime gradually reduces the source-resolving capabilities of microlensing and eventually suppresses microlensing in general by averaging over larger areas of the caustic network.

\section{SUMMARY}
\label{sec:summary}

In this work we used a simple model of a thin accretion disk around a Kerr black hole to study variations in a quasar X-ray spectrum during the crossing of a microlensing caustic. We demonstrated the gradual sequence of changes, especially the appearance and disappearance of additional features in the profile of the iron K$\alpha$ line. The energies of these peaks and edges indicated their origin from positions on the disk, at which the caustic was tangent to the $g$-factor (energy shift) contours.

To prove this indication, we employed a local approximation of the $g$-factor contours and showed that peaks were formed when a contour was tangent from the inner side of the caustic, while edges were formed when a contour was tangent from its outer side. We proceeded to derive the analytical shapes of these features and found that the peak had a logarithmic profile while the edge had a step function profile. We mapped the strength of the generated peak as a function of position of the tangent point on the disk. Maximum peak strength is achieved close to $g$-factor contour inflection points and near the $g$-factor maximum. We demonstrated the validity of the logarithmic approximation of the generated peak by performing fits to simulated microlensing spectra.

We discussed the limitations of the model and its possible extensions. We argued that --- due to the simple geometric origin of the generated line profile features --- the overall behavior should be generic, valid for a much broader range of quasar X-ray emission models. The sequence of variations in the generated line profile features then puts strong constraints on the geometry and physics of the emission region.

\acknowledgements

We thank Michal Bursa for providing us with his geodesic integration code. This work was supported by Czech Science Foundation grant GACR 14-37086G (Albert Einstein Center for Gravitation and Astrophysics) and by Charles University grant GAUK 2000314. MD thanks for the support through project LTAUSA17095 funded by MEYS, Czech Republic.

\clearpage
\bfi
\plotone{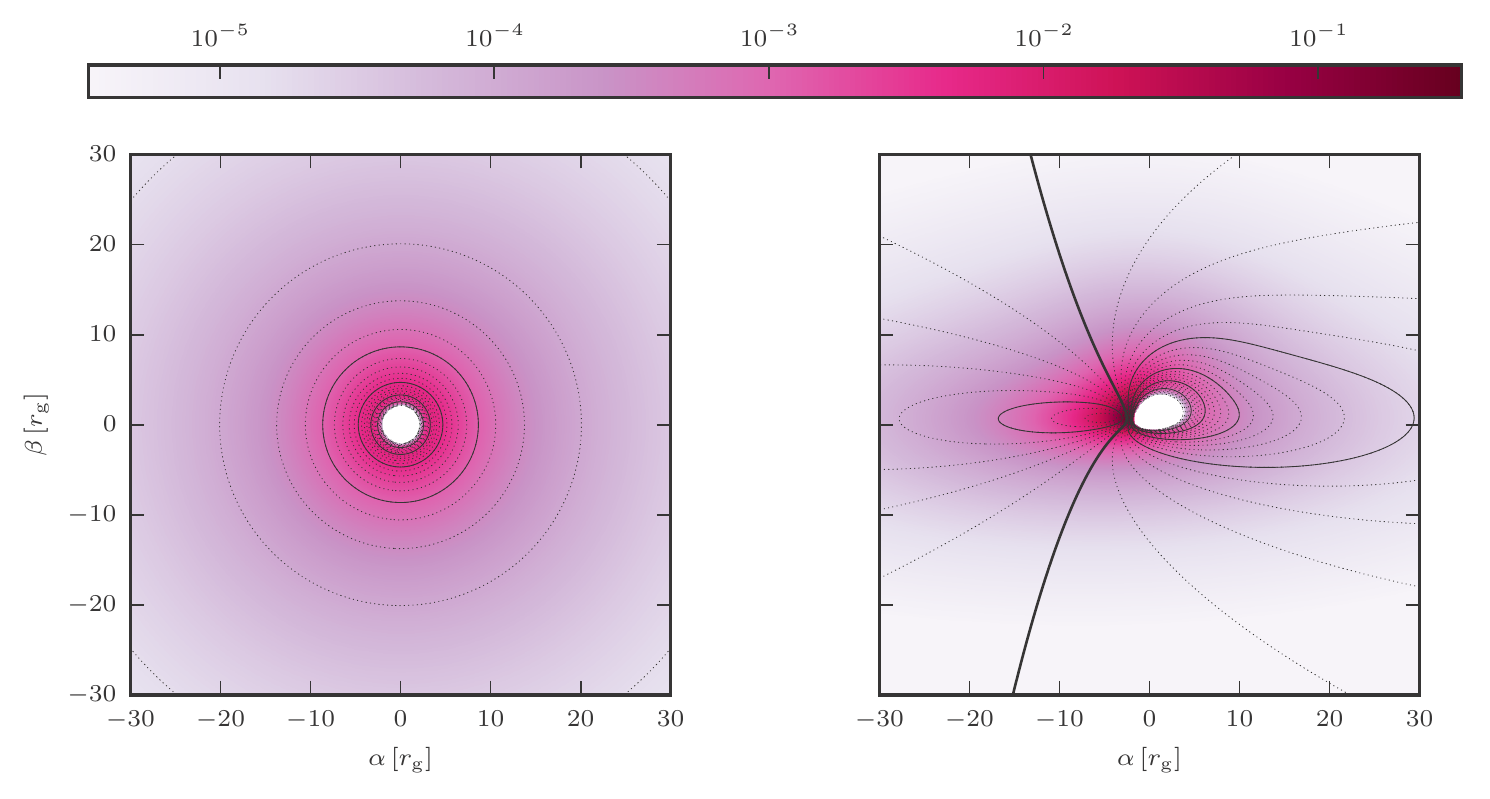}
\caption{Observer-frame maps of Fe K$\alpha$ line specific intensity $I_{\rm obs}^{\rm Fe}$ in the central part of the accretion disk around a Kerr black hole (spin $a=1$), shown for inclinations $i=0^\circ$ (left panel) and $i=70^\circ$ (right panel). Plane-of-the-sky coordinates $(\alpha,\beta)$ centered on the black hole are marked in gravitational radii $r_{\rm g}$. The observer-frame specific intensity can be described by a delta function in energy with position-dependent amplitude and energy shift, as given by Equation~(\ref{eq:intensity-iron-obs}). The amplitude in units of $I_0^{\rm Fe}\,r_{\rm g}^{-q}$ is mapped by color (for emissivity index $q=3$), while the energy shift is illustrated by $g$-factor contours: bold for $g=1.00$, dotted with step~$0.04$, solid with step~$0.20$. Left panel: contours decrease from outermost $g=0.96$ to innermost $g=0$ at horizon. Right panel: contours decrease from $g=1.28$ around point (-4.7, 0.7), via the top-to-bottom $g=1$, to $g=0$ at horizon.}
\label{fig:spec-int}
\efi

\clearpage
\bfi
{\centering \includegraphics[scale=1.29]{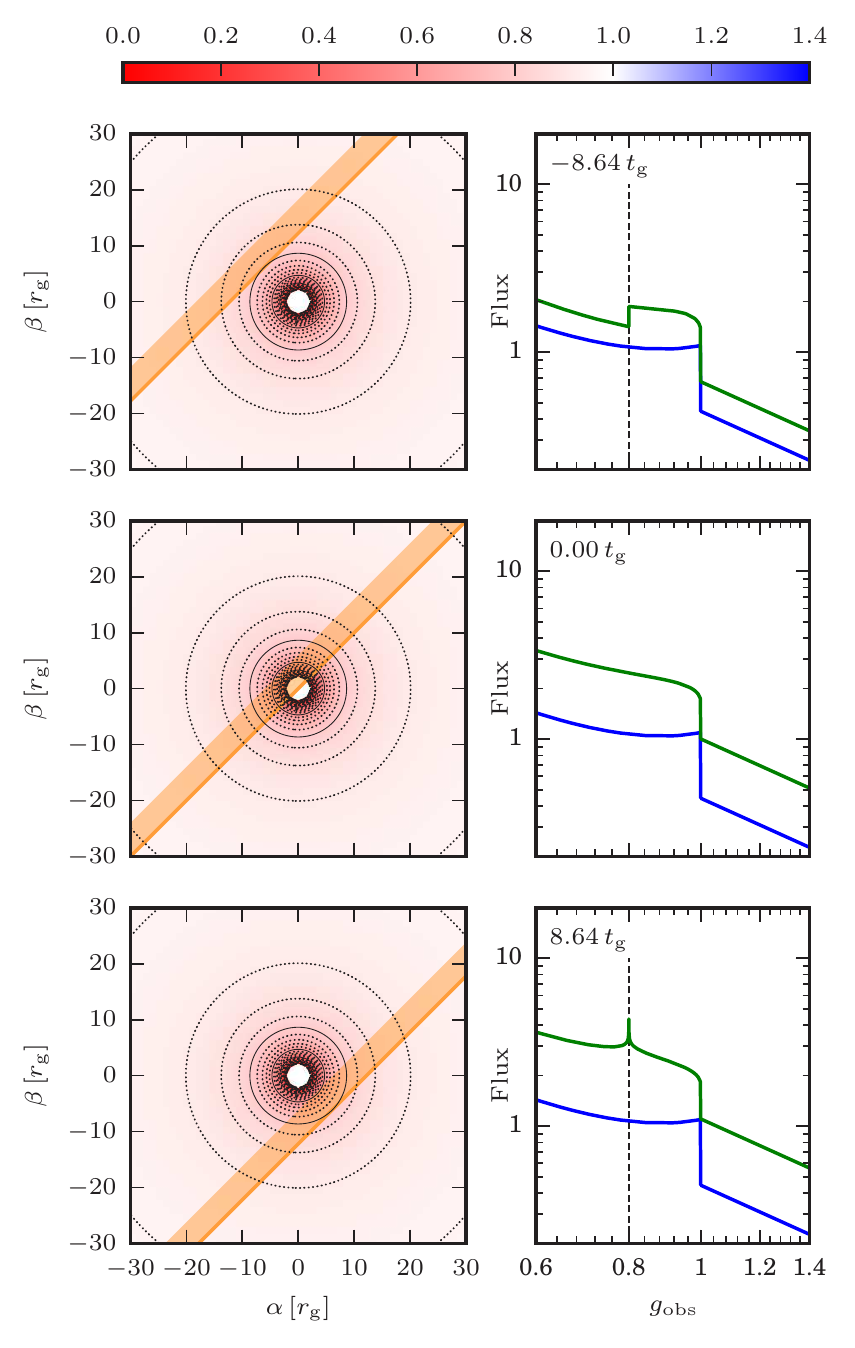}\par}
\caption{Fe K$\alpha$ line region in the X-ray spectrum of a face-on quasar accretion disk: three snapshots in the course of a microlensing event. Left panels: caustic position (dark-orange line; light-orange band indicates amplified side) plotted over color-mapped $g$-factor $g(\alpha,\beta)$ with the same contours as in Figure~\ref{fig:spec-int} (left panel). Right panels: total photon flux $F^{\rm cont}+F^{\rm Fe}$ in arbitrary units vs. $g_{\rm obs}$, photon energy in units of $E_{\rm Fe}$, showing microlensed spectrum (green curve) and spectrum in absence of microlensing (blue curve). Inscribed time $t$ is numerically equal to the distance of the caustic from the origin in $r_{\rm g}$ units. Dashed lines mark positions of generated edge (top row) and peak (bottom row). For more details see \S~\ref{sec:evol-0}.}
\label{fig:spec-evol-0}
\efi

\clearpage
\bfi
{\centering \includegraphics[scale=1.14]{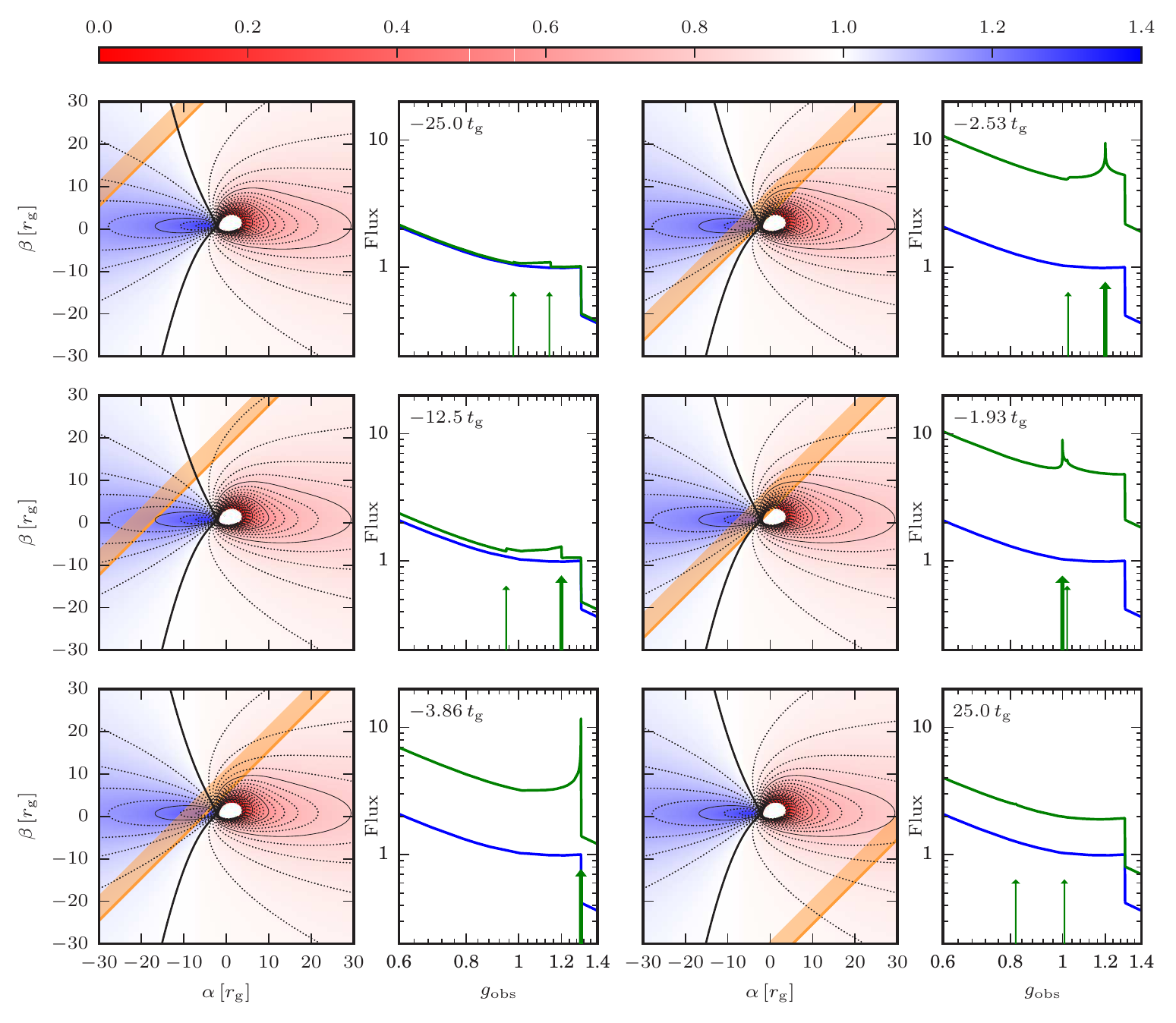}\par}
\caption{Fe K$\alpha$ line region in the X-ray spectrum of an $i=70^\circ$ inclined quasar accretion disk: six snapshots in the course of a microlensing event. Notation is the same as in Figure~\ref{fig:spec-evol-0}. $G$-factor contours are the same as in Figure~\ref{fig:spec-int} (right panel). Green arrows mark positions of generated features (major features in bold). For more details see \S~\ref{sec:evol-70}.}
\label{fig:spec-evol-70}
\efi

\clearpage
\bfi
{\centering \includegraphics[scale=1.29]{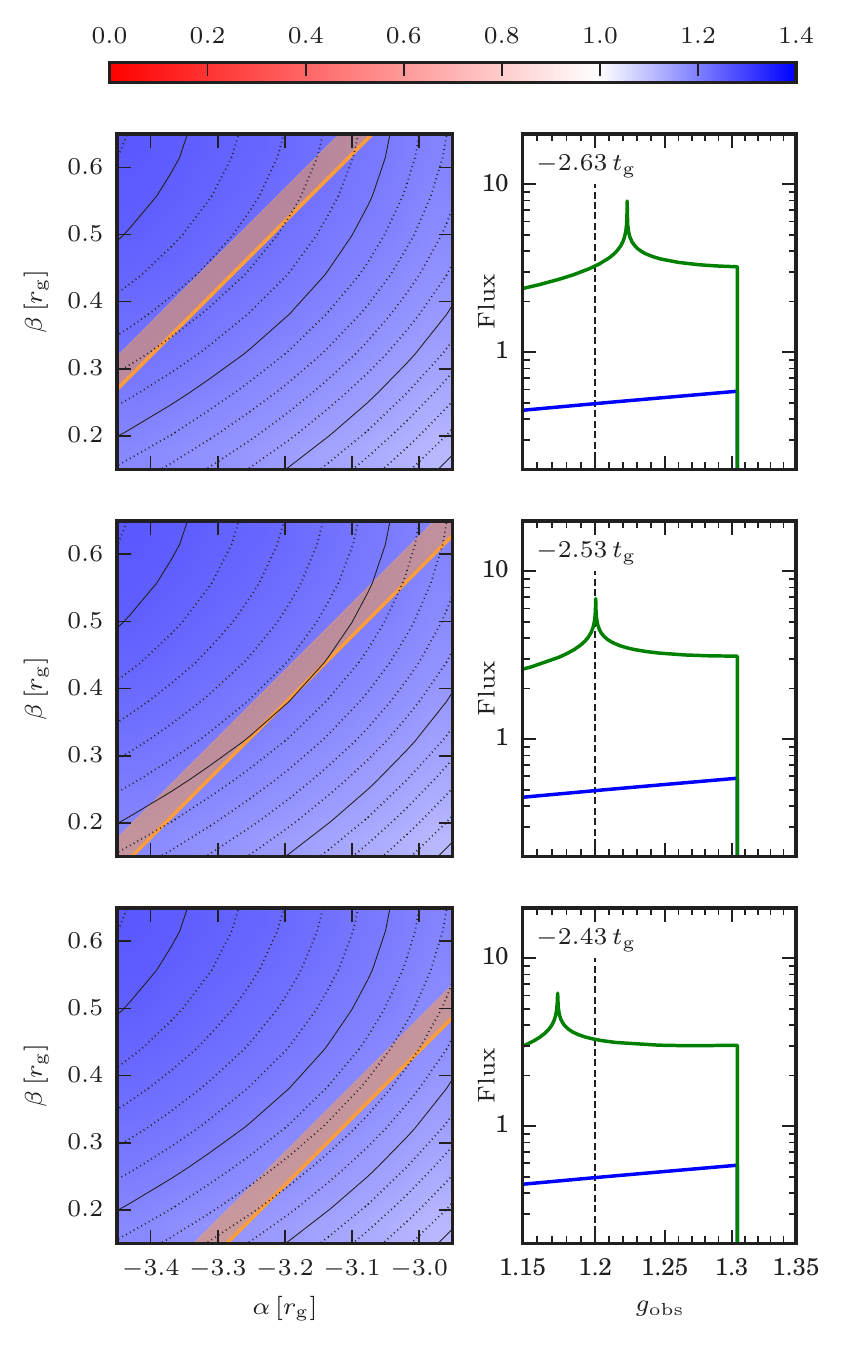}\par}
\caption{Evolution of a generated peak in the Fe K$\alpha$ line profile: three snapshots around the configuration from the top right panel of Figure~\ref{fig:spec-evol-70}. Left panels: detail of $g$-factor map near the tangent point of the $g=1.20$ contour and caustic at $t=-2.53\,t_{\rm g}$. Contour spacing is finer here: dotted with step~$0.01$, solid with step~$0.05$. Right panels: detail of line profile showing only line flux $F^{\rm Fe}$. Dashed lines mark $g_{\rm obs}=1.20$ peak position at $t=-2.53\,t_{\rm g}$. The remaining notation is the same as in Figure~\ref{fig:spec-evol-0}. For more details see \S~\ref{sec:featuresPE}.}
\label{fig:spec-evol-peak}
\efi

\clearpage
\bfi
{\centering \includegraphics[scale=1.29]{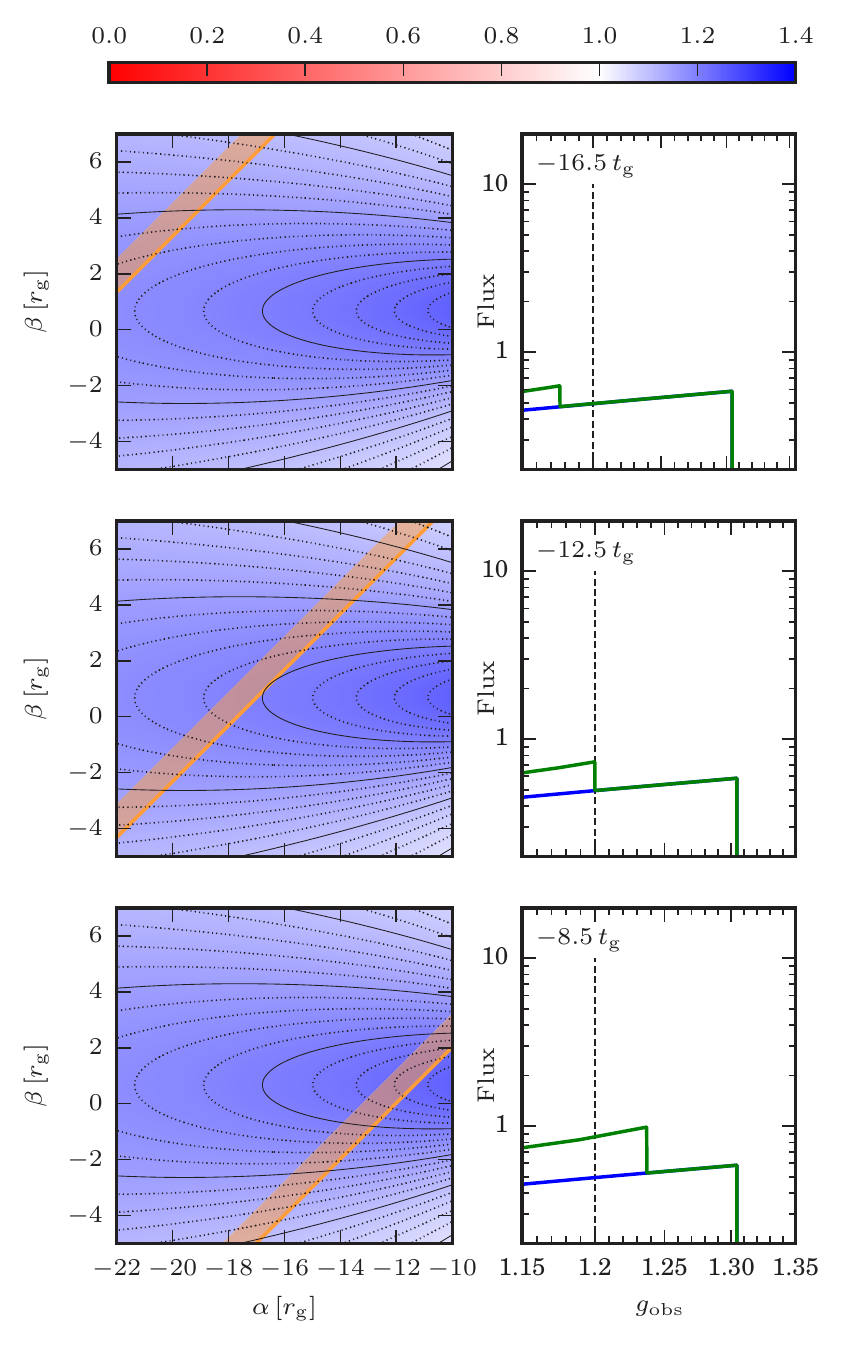}\par}
\caption{Evolution of a generated edge in the Fe K$\alpha$ line profile: three snapshots around the configuration from the middle left panel of Figure~\ref{fig:spec-evol-70}. Left panels: detail of $g$-factor map near the tangent point of the $g=1.20$ contour and caustic at $t=-12.5\,t_{\rm g}$. Dashed lines in the right panels mark the $g_{\rm obs}=1.20$ edge position at $t=-12.5\,t_{\rm g}$. The remaining notation is the same as in Figure~\ref{fig:spec-evol-peak}. For more details see \S~\ref{sec:featuresPE}.}
\label{fig:spec-evol-edge}
\efi

\clearpage
\bfi
{\centering \includegraphics[scale=1.29]{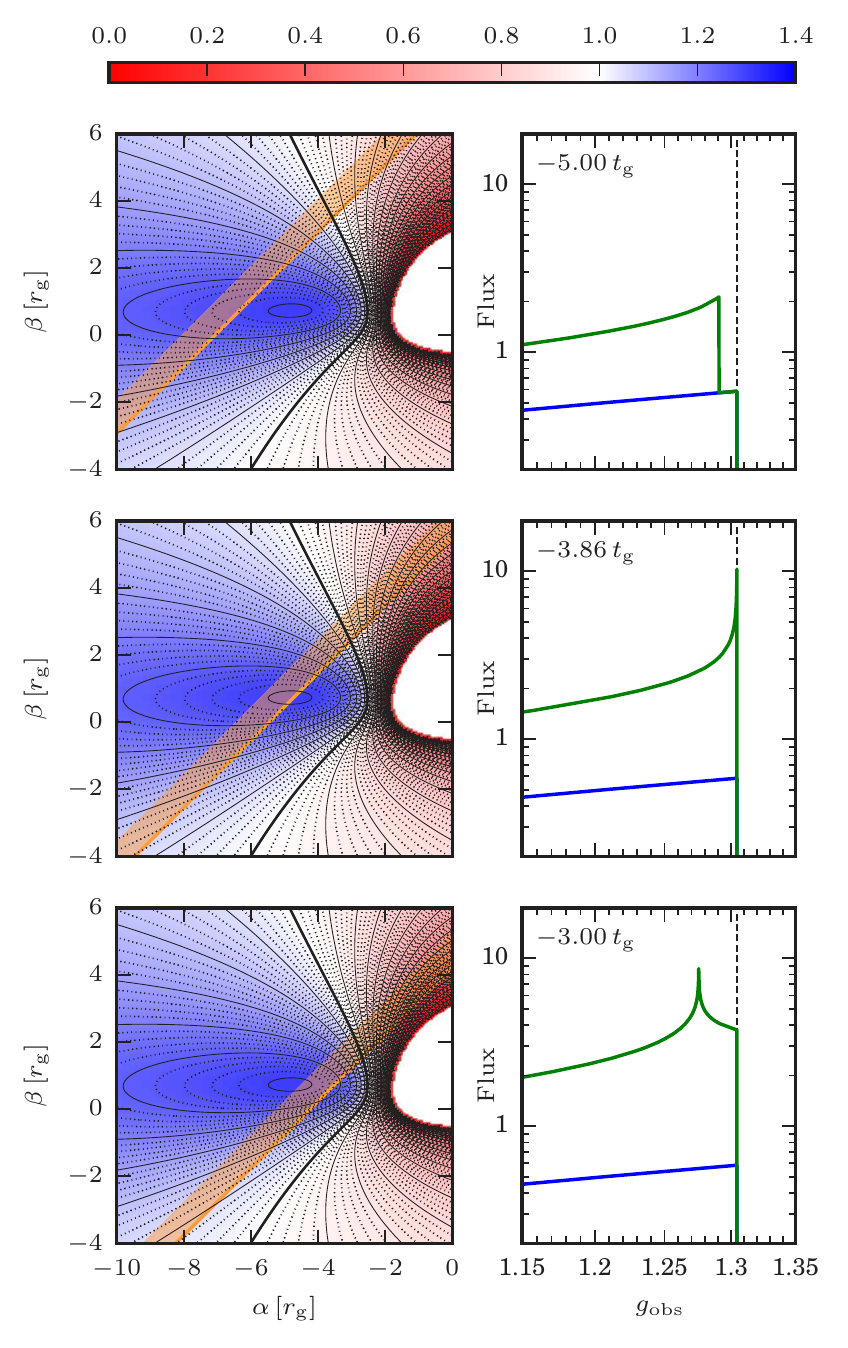}\par}
\caption{Evolution of the Fe K$\alpha$ line profile: three snapshots around $g$-factor maximum crossing from the bottom left panel of Figure~\ref{fig:spec-evol-70}. Left panels: detail of $g$-factor map near $g$-factor maximum $g_{\rm max}=1.304$. Dashed lines in right panels mark the $g_{\rm obs}=1.304$ asymmetric peak position at $t=-3.86\,t_{\rm g}$. The remaining notation is the same as in Figure~\ref{fig:spec-evol-peak}. For more details see \S~\ref{sec:specf-higher}.}
\label{fig:spec-evol-max}
\efi

\clearpage
\bfi
{\centering \includegraphics[scale=1.29]{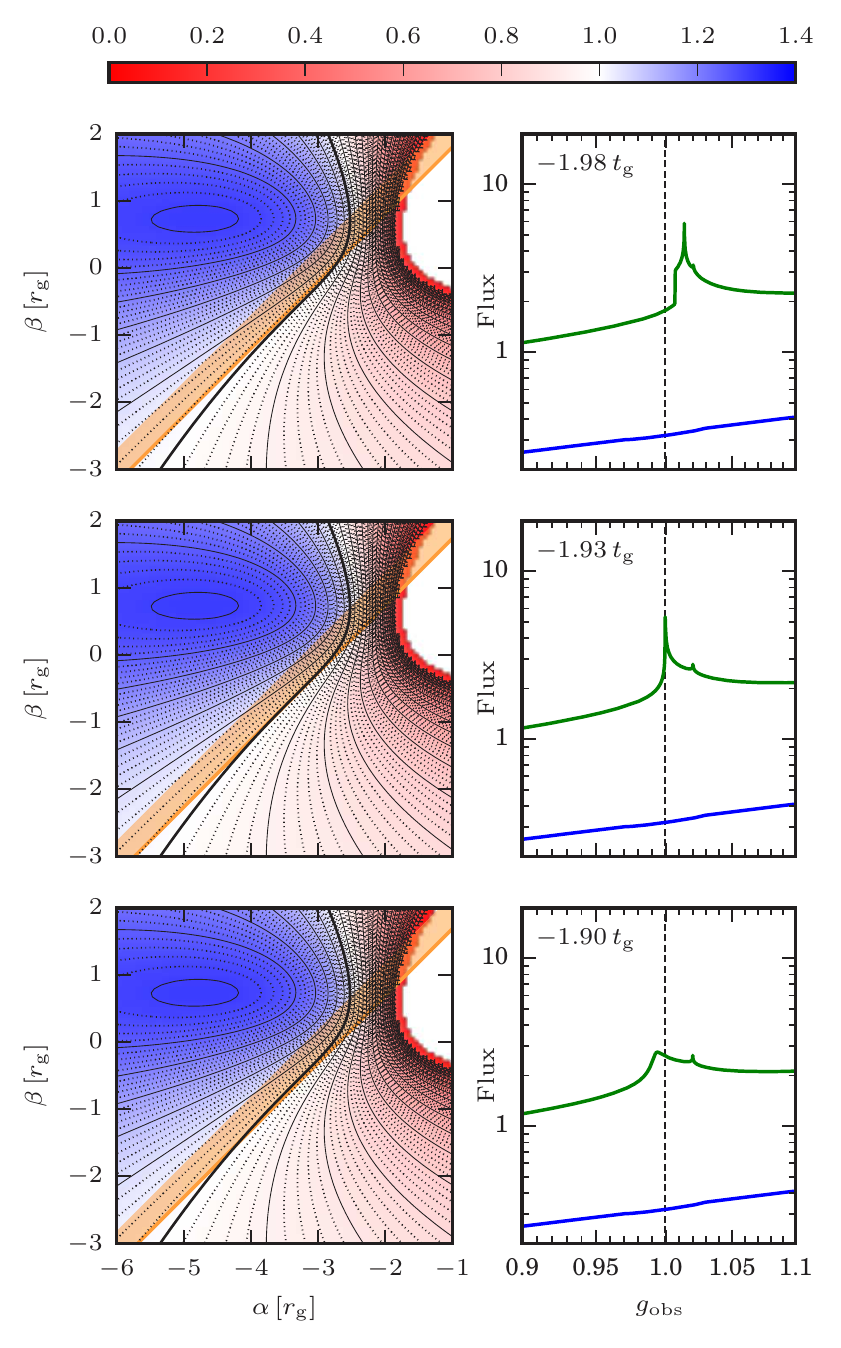}\par}
\caption{Evolution of the Fe K$\alpha$ line profile: three snapshots around the tangent crossing of the $g$-factor contour inflection point from the middle right panel of Figure~\ref{fig:spec-evol-70}. Left panels: detail of $g$-factor map near the tangent point of the $g=0.9996$ contour and the caustic at $t=-1.93\,t_{\rm g}$. Dashed lines in right panels mark the $g_{\rm obs}=0.9996$ asymmetric peak position at $t=-1.93\,t_{\rm g}$. The remaining notation is the same as in Figure~\ref{fig:spec-evol-peak}. For more details see \S~\ref{sec:specf-higher}.}
\label{fig:spec-evol-infl}
\efi

\clearpage
\bfi
\begin{center}
\includegraphics{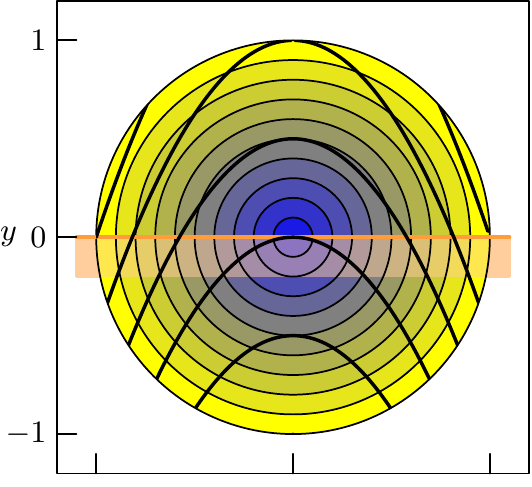}
\hspace{3mm}
\includegraphics{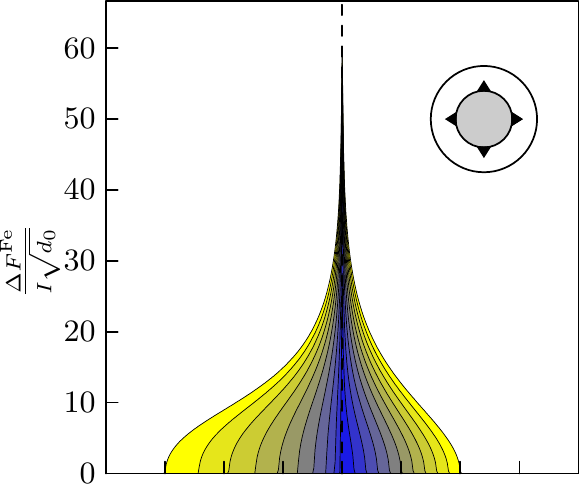}
\includegraphics{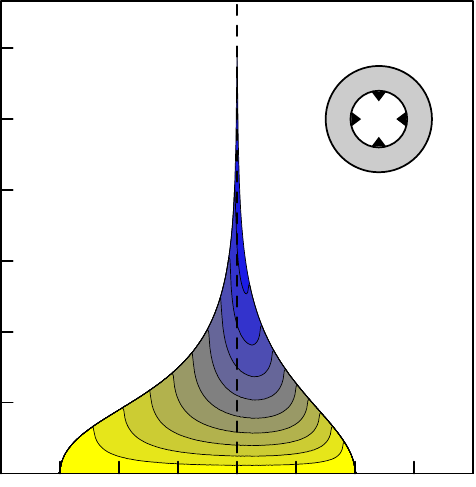}\\
\includegraphics{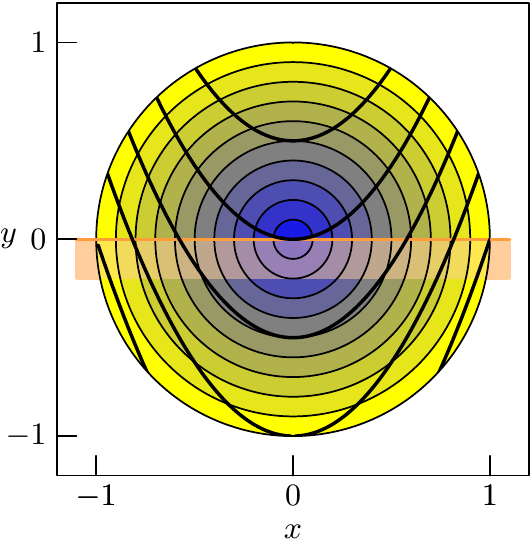}
\hspace{3mm}
\includegraphics{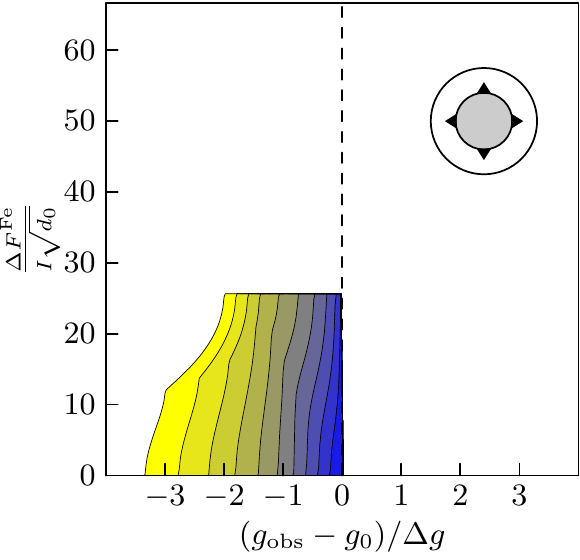}
\includegraphics{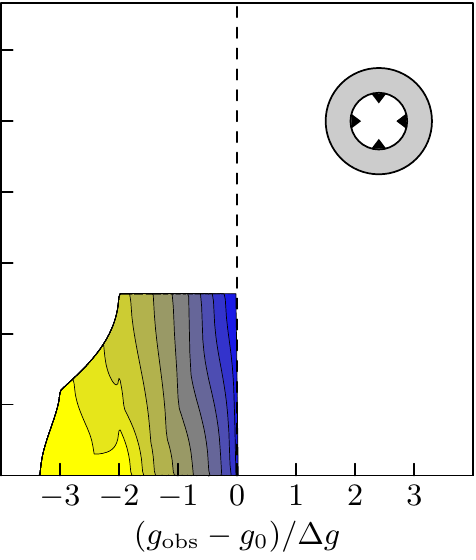}
\end{center}
\caption{Excess flux spectra generated by the caustic and a circular source, detailing contributions to the line profile from different regions of the source. Top row: $g_0$ contour tangent from inside the caustic demonstrating peak generation; bottom row: $g_0$ contour tangent from outside demonstrating edge generation. Left panels: parabolic lines --- $g$-factor contours with $g_0$ passing through center; orange line -- caustic with inner side indicated by band; colors from blue to yellow mark annular parts of the source contributing to the line profiles. Top row, $g_y=-0.1$ ($g$-factor decreasing upward); bottom row, $g_y=0.1$ ($g$-factor increasing upward); both rows, $g_{xx}=-0.3$. Middle panels: spectrum of a circular part of the source with outer radius increasing from blue to yellow. Right panels: spectrum of an annular part of the source with inner radius decreasing from yellow to blue. Spectral width parameter $\Delta g=0.05$.}
\label{fig:circ-model-size}
\efi

\clearpage
\bfi
\begin{center}
\includegraphics{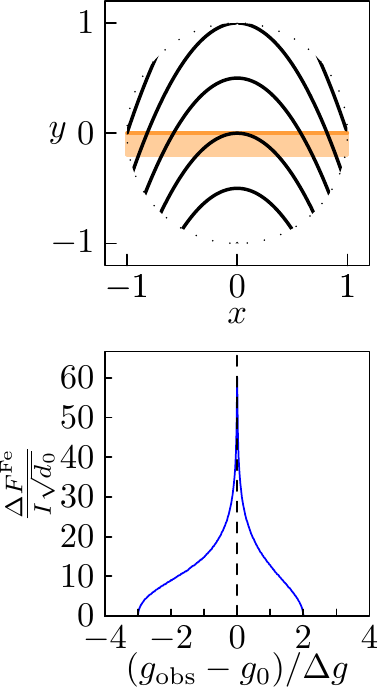}
\includegraphics{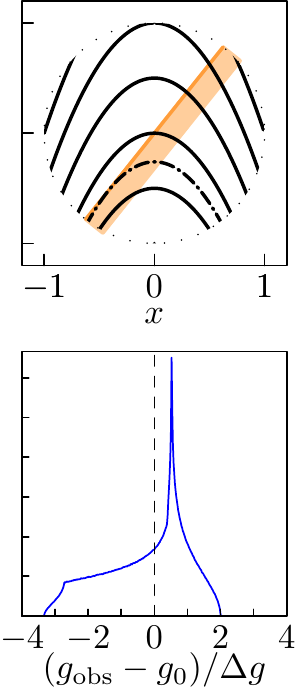}
\includegraphics{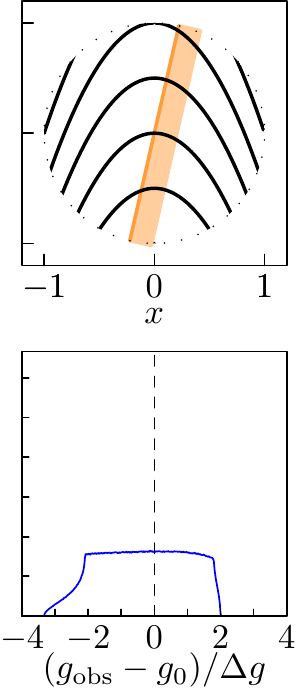}
\includegraphics{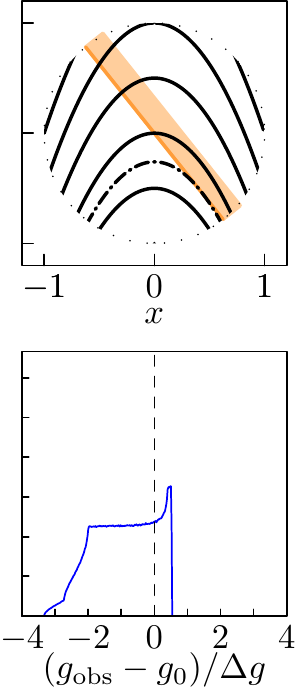}
\includegraphics{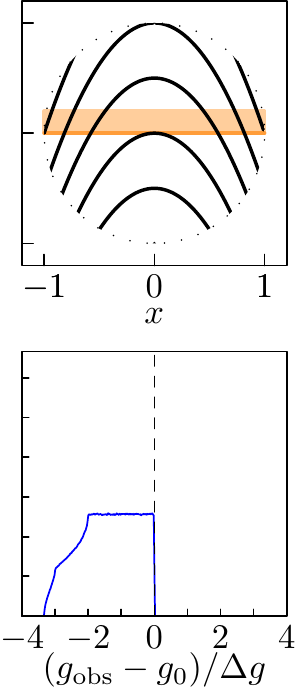}
\end{center}
\caption{Excess flux spectra generated by the caustic and a circular source, plotted for different caustic orientations. Top row: $g$-factor contours with $g=g_0$ passing through center and $g$ decreasing upward ($g_y=-0.1$, $g_{xx}=-0.3$, step $\Delta g = 0.05$); tangent contours in the second and fourth columns are marked by dot-dashed lines. Bottom row: corresponding spectra with $g_0$ marked by dashed lines. The rest of the notation is the same as in Figure~\ref{fig:circ-model-size}.}
\label{fig:circ-model}
\efi

\clearpage
\bfi
\plotone{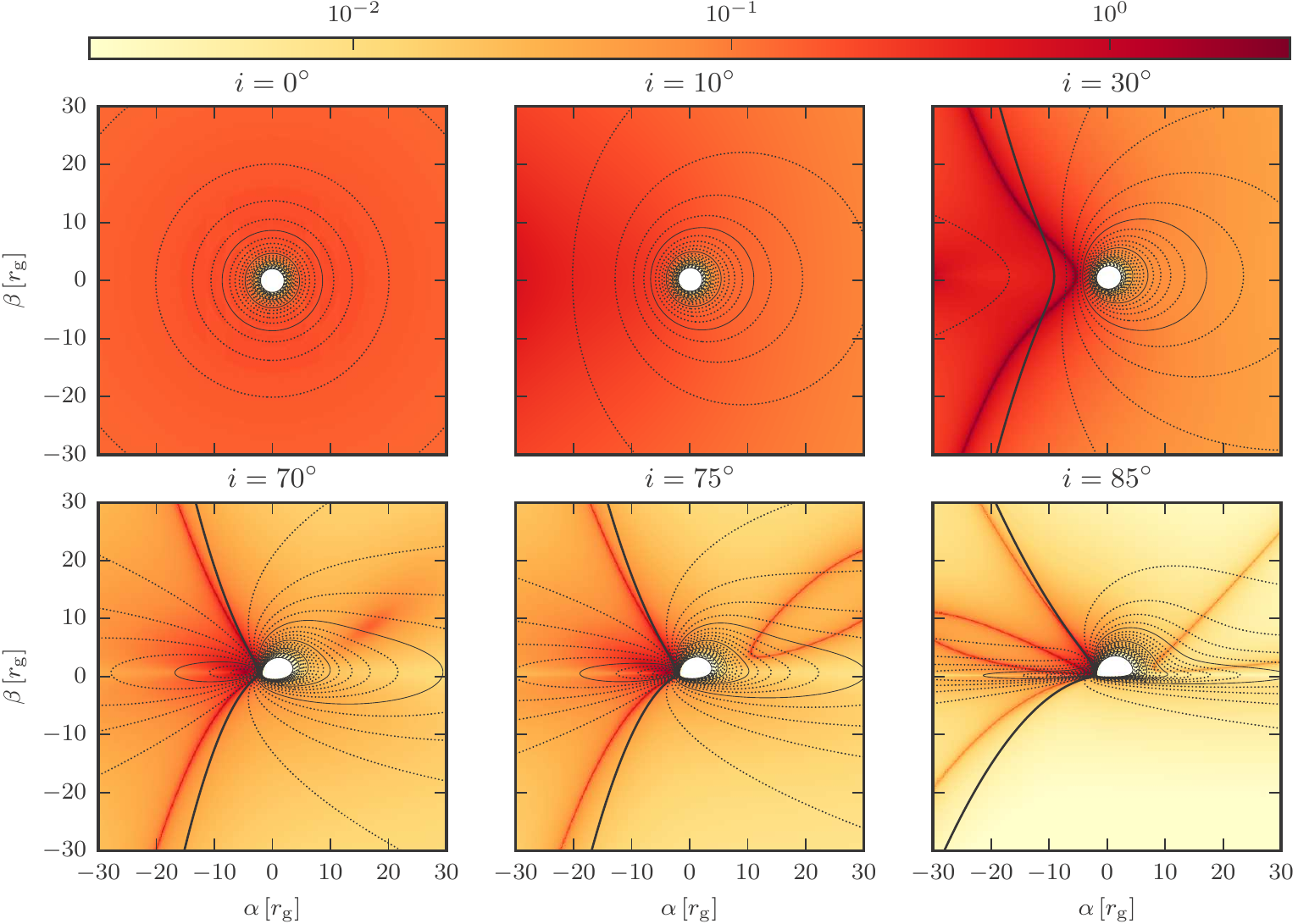}
\caption{Maps of the peak strength $P$ given by Equation~(\ref{eq:peak-strng}) indicating the prominence of generated peaks or edges, for several disk inclinations marked above the panels. For a given point in a plot, a peak in the line profile is generated if the caustic passes through the point oriented so that the local $g$-factor contour is tangent from the amplified side, while an edge is generated if the contour is tangent from the unamplified side. $G$-factor contours are plotted from $g = 0$ at horizon: dotted with step $0.04$, solid with step $0.20$, bold for $g = 1.00$. Peak strength $P$ is indicated by the color bar in units of $F_0^{\rm Fe}\,r_{\rm g}^{-q}$.}
\label{fig:peak-strg}
\efi

\clearpage
\bfi
\includegraphics[scale=1.3]{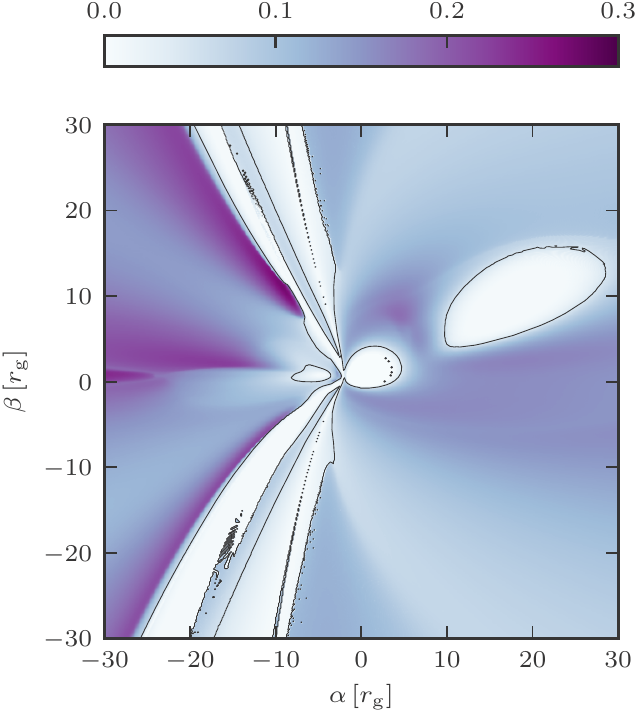}
\hspace{5mm}
\includegraphics[scale=1.3]{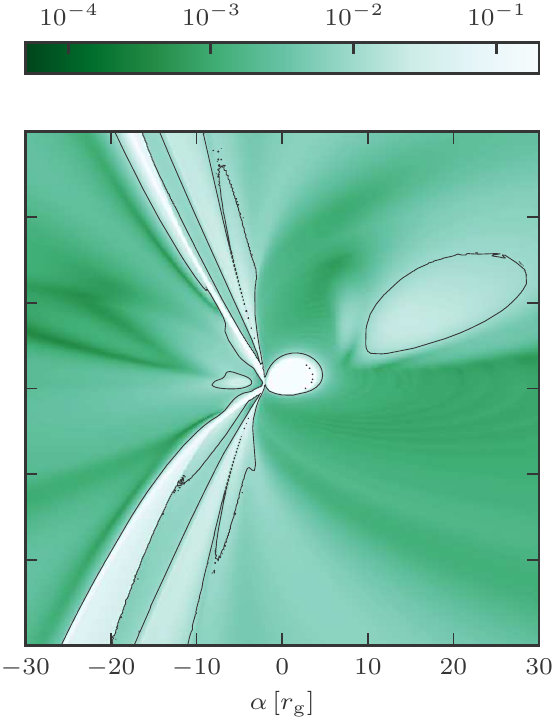}
\caption{Accuracy of the microlensing-generated peak model from Equation~(\ref{eq:peak-model}), computed for disk inclination $70^\circ$. Left panel: map of maximum spectral interval half-width $\Delta g$ around the peak maintaining relative error $\epsilon \leq 1\%$, with the $\Delta g=0.05$ contour plotted. Right panel: map of relative error $\epsilon$ for fixed half-width $\Delta g=0.05$, with $\epsilon = 1\%$ contour plotted.}
\label{fig:peak-accuracy}
\efi


\begin{thebibliography}{}

\bibitem[Abolmasov \& Shakura(2012)]{abolmasov_shakura12}
Abolmasov, P., \& Shakura, N.~I.\ 2012, \mnras, 423, 676

\bibitem[Agol \& Krolik(1999)]{agol_etal99}
Agol, E., \& Krolik, J.\ 1999, \apj, 524, 49

\bibitem[Beckwith \& Done(2004)]{beckwith_done04}
Beckwith, K., \& Done, C.\ 2004, \mnras, 352, 353

\bibitem[Beckwith \& Done(2005)]{beckwith_done05}
Beckwith, K., \& Done, C.\ 2005, \mnras, 359, 1217

\bibitem[Chartas et al.(2002)]{chartas_etal02}
Chartas, G., Agol, E., Eracleous, M., et al.\ 2002, \apj, 568, 509

\bibitem[Chartas et al.(2012)]{chartas_etal12}
Chartas, G., Kochanek, C.~S., Dai, X., et al.\ 2012, \apj, 757, 137

\bibitem[Chartas et al.(2009)]{chartas_etal09}
Chartas, G., Kochanek, C.~S., Dai, X., Poindexter, S., \& Garmire, G.\ 2009, \apj, 693, 174

\bibitem[Chartas et al.(2017)]{chartas_etal17}
Chartas, G., Krawczynski, H., Zalesky, L., et al.\ 2017, \apj, 837, 26

\bibitem[Chen et al.(2013a)]{chen_etal13a}
Chen, B., Dai, X., \& Baron, E.\ 2013a, \apj, 762, 122

\bibitem[Chen et al.(2013b)]{chen_etal13b}
Chen, B., Dai, X., Baron, E., \& Kantowski, R.\ 2013b, \apj, 769, 131

\bibitem[Dai et al.(2010)]{dai_etal10}
Dai, X., Kochanek, C.~S., Chartas, G., et al.\ 2010, \apj, 709, 278

\bibitem[Dov{\v c}iak et al.(2004)]{dovciak_etal04}
Dov{\v c}iak, M., Karas, V., \& Yaqoob, T.\ 2004, \apjs, 153, 205

\bibitem[Eigenbrod et al.(2008)]{eigenbrod_etal08}
Eigenbrod, A., Courbin, F., Meylan, G., et al.\ 2008, \aap, 490, 933

\bibitem[Fabian(2006)]{fabian06}
Fabian, A.~C.\ 2006, Astronomische Nachrichten, 327, 943

\bibitem[Fukue \& Akizuki(2006)]{fukue_akizuki06}
Fukue, J., \& Akizuki, C.\ 2006, \pasj, 58, 1073

\bibitem[Guerras et al.(2017)]{guerras_etal17}
Guerras, E., Dai, X., Steele, S., et al.\ 2017, \apj, 836, 206

\bibitem[Heyrovsk{\'y} \& Loeb(1997)]{heyrovsky_etal97}
Heyrovsk{\'y}, D., \& Loeb, A.\ 1997, \apj, 490, 38

\bibitem[Jaroszy{\'n}ski et al.(1992)]{jaroszynski_etal92}
Jaroszy\'nski, M., Wambsganss, J., \& Paczy\'nski, B.\ 1992, \apjl, 396, L65

\bibitem[Jim{\'e}nez-Vicente et al.(2015)]{jimenez-vicente_etal15}
Jim{\'e}nez-Vicente, J., Mediavilla, E., Kochanek, C.~S., \& Mu{\~n}oz, J.~A.\ 2015, \apj, 806, 251

\bibitem[Jovanovi{\'c}(2012)]{jovanovic12}
Jovanovi{\'c}, P.\ 2012, \nar, 56, 37

\bibitem[Jovanovi{\'c} et al.(2009)]{jovanovic_etal09}
Jovanovi{\'c}, P., Popovi{\'c}, L.~{\v C}., \& Simi{\'c}, S.\ 2009, \nar, 53, 156

\bibitem[Krawczynski \& Chartas(2017)]{krawczynski_chartas17}
Krawczynski, H., \& Chartas, G.\ 2017, \apj, 843, 118

\bibitem[Laor(1991)]{laor91}
Laor, A.\ 1991, \apj, 376, 90

\bibitem[Ledvina \& Heyrovsk\'y(2015)]{ledvina_heyrovsky15}
Ledvina, L., \& Heyrovsk\'y, D. 2015, in WDS'15 Proc. Contributed Papers -- Physics, ed. J.~\v{S}afr\'ankov\'a \& J.~Pavl\r{u} (Prague: Matfyzpress), 21, \url{https://www.mff.cuni.cz/veda/konference/wds/proc/pdf15/WDS15_03_f1_Ledvina.pdf}

\bibitem[MacLeod et al.(2015)]{macleod_etal15}
MacLeod, C.~L., Morgan, C.~W., Mosquera, A., et al.\ 2015, \apj, 806, 258

\bibitem[Mediavilla et al.(2015)]{mediavilla_etal15}
Mediavilla, E., Jim{\'e}nez-Vicente, J., Mu{\~n}oz, J.~A., \& Mediavilla, T.\ 2015, \apjl, 814, L26

\bibitem[Morgan et al.(2012)]{morgan_etal12}
Morgan, C.~W., Hainline, L.~J., Chen, B., et al.\ 2012, \apj, 756, 52

\bibitem[Mu{\~n}oz et al.(2016)]{munoz_etal16}
Mu{\~n}oz, J.~A., Vives-Arias, H., Mosquera, A.~M., et al.\ 2016, \apj, 817, 155

\bibitem[Neronov \& Vovk(2016)]{neronov_etal16}
Neronov, A., \& Vovk, I.\ 2016, \prd, 93, 023006

\bibitem[Popovi{\'c} et al.(2006)]{popovic_etal06}
Popovi{\'c}, L.~{\v C}., Jovanovi{\'c}, P., Mediavilla, E., et al.\ 2006, \apj, 637, 620

\bibitem[Popovi{\'c} et al.(2003a)]{popovic_etal03a}
Popovi{\'c}, L.~{\v C}, Jovanovi{\'c}, P., Mediavilla, E., \& Mu{\~n}oz, J.~A.\ 2003a, A\&AT, 22, 719

\bibitem[Popovi{\'c} et al.(2003b)]{popovic_etal03b}
Popovi{\'c}, L.~{\v C}., Mediavilla, E.~G., Jovanovi{\'c}, P., \& Mu{\~n}oz, J.~A.\ 2003b, \aap, 398, 975

\bibitem[Popovi{\'c} et al.(2001)]{popovic_etal01}
Popovi{\'c}, L.~{\v C}, Mediavilla, E.~G., \& Mu{\~n}oz, J.~A.\ 2001, \aap, 378, 295

\bibitem[Schmidt \& Wambsganss(2010)]{schmidt_etal10}
Schmidt, R.~W., \& Wambsganss, J.\ 2010, General Relativity and Gravitation, 42, 2127

\bibitem[Schneider et al.(1992)]{schneider_etal92}
Schneider, P., Ehlers, J., \& Falco, E.~E.\ 1992, Gravitational Lenses (Berlin: Springer)

\bibitem[Walton et al.(2015)]{walton_etal15}
Walton, D.~J., Reynolds, M.~T., Miller, J.~M., et al.\ 2015, \apj, 805, 161

\bibitem[Yonehara et al.(1998)]{yonehara_etal98}
Yonehara, A., Mineshige, S., Manmoto, T., et al.\ 1998, \apjl, 501, L41

\bibitem[Zimmer et al.(2011)]{zimmer_etal11}
Zimmer, F., Schmidt, R.~W., \& Wambsganss, J.\ 2011, \mnras, 413, 1099

\end{thebibliography}
\end{document}